\documentclass{IEEEtran}
\usepackage{cite}
\usepackage{amsmath,amssymb,amsfonts}
\usepackage{algorithmic}
\usepackage{subcaption}
\usepackage{graphicx}
\usepackage{textcomp}
\def\BibTeX{{\rm B\kern-.05em{\sc i\kern-.025em b}\kern-.08em
    T\kern-.1667em\lower.7ex\hbox{E}\kern-.125emX}}
\begin{document}
\title{Accuracy and Efficiency of Lowest-order and Higher-Order Single-Source Integral Equations}
\author{A. Zuccotti, \IEEEmembership{Student Member, IEEE}, C. M{\"u}nger, \IEEEmembership{Member, IEEE}, and K. Cools, \IEEEmembership{Member, IEEE}

\thanks{This work was supported by European Research Council (ERC) under European Union’s Horizon 2020 Research and Innovation Program under Grant 101001847 and by Special Research Fund (BOF) of Ghent University under Grant BOF.PDO.2024.0016.01 .}
\thanks{A. Zuccotti, K. Cools, and C. M\"{u}nger are with IDLab, Department of Information Technology at Ghent University - imec, Ghent, Belgium (Corresponding author: Alessandro Zuccotti e-mail: alessandro.zuccotti@ugent.be).}
\thanks{This work has been submitted to the IEEE for possible publication. Copyright may be transferred without notice, after which this version may no longer be accessible.}}

\maketitle 

\begin{abstract}
The single-source integral equation allows the modeling of the electromagnetic field by only tracking a single surface-supported current density. It features products of integral operators, which pose unique challenges for the modeler in both the discretization and solution of the equation.
Due to the unique structure of the single-source equation, its conforming and stable discretization requires expanding one of the two physical traces in primal basis functions and the other in dual basis functions. Primal functions are expected to have better approximation properties than dual ones. In this work, we study the impact of this asymmetric representation on the accuracy of the solution and propose an approach from which both traces can be recovered using only primal functions with modest additional computational cost.

Moreover, we propose an alternative discretization scheme based on approximating the inverse of the duality mapping induced by the surface normal cross-product operator through the quasi-Helmholtz decomposition.
This discretization provides greater flexibility in the choice of function spaces, enabling both a higher-order extension of the single-source equation and a more efficient lowest-order discretization using only Rao-Wilton-Glisson basis functions. 

Numerical results show the impact of the representation in terms of primal and dual functions in both lowest- and higher-order discretization. 
\end{abstract}

\begin{IEEEkeywords}
Boundary element methods, Buffa-Christiansen functions, dielectrics, dual space, higher-order,  homogeneous object, integral equations, preconditioning, quasi-Helmholtz projectors, single-source equation.

\end{IEEEkeywords}
\section{Introduction}
\IEEEPARstart{B}{oundary} integral equations (BIEs) provide an efficient and accurate framework for modeling electromagnetic scattering from both perfect electric conductors and dielectric materials. For homogeneous dielectric problems, the most widely used formulations are the PMCHWT and Müller equations \cite{poggio1970integral, Chang_Harrington_PMCHWT, Wu_Tsai_PMCHWT, muller2013foundations}, which have been extensively investigated over the past decades.
These equations are derived by adding or subtracting the internal and external Calderón formulas for the corresponding electric and magnetic traces on the boundary of the scatterer. Both formulations yield a coupled system for the electric and magnetic equivalent surface currents.

In addition to these classical formulations, single-source integral equations were originally introduced in \cite{Marx1984, Glisson1984,costabel2011kleinman}.
Single-source equations are constructed by combining the Calderón formulas of the exterior and interior problems in a multiplicative manner. The interior problem admits non-unique solutions, and this redundancy allows one to select an ansatz that leads to one of a number of possible single-source formulations.
The resulting system involves only one auxiliary current, either of electric or magnetic type, from which both physical surface traces can be recovered. The main advantage is that the number of unknowns is reduced by half compared to the PMCHWT or Müller equations. Moreover, the single-source formulation can be made intrinsically well conditioned, and thus immune to dense-mesh breakdown by properly selecting the ansatz, without the need for constructing a preconditioner or performing the extra matrix-vector products required for its inclusion in the iterative solution process. The single-source equation, when used in the time domain, has advantages in terms of late-time stability. Promising results have been presented in \cite{ValdesTDSingleSource}. It is expected that the stabilization methods used in \cite{CoolsAndriulliTDCalderonPart2} and \cite{Chien2024TDCFIE} can lead to even more robust implementations.

However, single-source formulations are typically affected by internal resonances. This issue can be mitigated by appropriately combining two different ansätze of the single-source equation \cite{Valdes2011}. The same combined approach can also be employed to suppress PEC resonances in composite configurations \cite{Munger2024}. Furthermore, the single-source formulation can be combined with a multi-trace approach for problems involving multiple domains and junctions. In this scenario, halving the number of unknowns is particularly advantageous \cite{Coolsmultitrace2021, Coolsmultitrace2022}.

Due to the presence of products of multiple integral operators, the discretization of the single-source equation requires both a primal space (e.g., the Rao–Wilton–Glisson space, RWG \cite{rao_electromagnetic_1982}) and a dual space defined on a mesh refinement (e.g., the Buffa–Christiansen space, BC, or Chen-Wilton space \cite{BUFFA2005461}, \cite{chen1990electromagnetic}). The conforming discretization in terms of RWG and BC basis functions ensures an inf-sup stable discretization of single- and double-layer boundary integral operators. Moreover, it results in the cancellation of low-frequency terms of order $O(1/\omega^2)$ in the multiplication of two single-layer operators. The single-source equation may exhibit numerical cancellation at very low frequencies, while high-contrast cases may also be challenging. These issues are not considered here and are left for future work.
In the conforming discretization, the auxiliary unknown is typically expanded in terms of primal RWG functions. However, the magnetic and electric traces are then obtained by applying the discrete internal single-layer and double-layer operators to the auxiliary solution. As a result, one of the two physical traces is expanded in terms of dual BC functions, while the other is expanded in terms of primal RWG functions. The expansion in terms of dual functions is expected to be less accurate than that in terms of primal functions, especially in the vicinity of sharp edges and corners of non-smooth geometries. In \cite{BUFFA2005461}, the approximation properties of the BC space are established via $H^{0}(\operatorname{div}; \Gamma)$ error estimate bounded by $O(h^{1/2})$, whereas $O(h)$ estimate can be reached for a sufficiently regular solution expanded in the RWG space \cite{christiansen2002}. Moreover, each BC basis functions represents a current that flows along an edge of the mesh and therefore introduce spurious continuity across sharp edges (being quasi curl-conforming), whereas solutions for the electric and magnetic traces in general exhibit discontinuities in their tangential components along lines where the surface normal changes in an abrupt manner.
This imbalance in accuracy between the two physical traces has not been addressed in the literature. The first aim of this work is to assess the severity of this limitation in accuracy and present strategies to mitigate it.

In this work we demonstrate that two different single-source ans\"atze can be derived from the representation formulas. With a suitable choice of discretization, both traces can be obtained in the primal space without requiring all integral operators to be discretized twice, hence within acceptable computational costs, although two linear systems still need to be solved.

The second aim of this work is to discuss a generalized discretization scheme that enables a higher-order extension of the single-source equation.

Again, primal and dual functions are required for the accurate and efficient discretization of the single-source equation. A popular higher-order generalization of the RWG basis functions is the Graglia–Wilton–Peterson (GWP) basis proposed in \cite{graglia_higher_1997}.

Past works have explored higher-order extensions of BC dual functions as well. In \cite{valdes_2011_higher_order}, higher-order dual functions are constructed by mapping the solenoidal (loop) and non-solenoidal (star) components of the basis functions into the rotated non-solenoidal and solenoidal components of the same basis functions, defined on the barycentric refinement.

In \cite{christiansen2002}, an alternative approach was used to precondition the EFIE. Instead of dual functions, general auxiliary div-conforming functions were employed to discretize the duality mapping between curl-conforming and div-conforming spaces induced by the surface normal cross-product operator, denoted by $\hat{\mathbf{n}} \times$. In particular, a general auxiliary space and the quasi-Helmholtz decomposition of the primal space into solenoidal and non-solenoidal components were used to define a map $\Theta$ that approximates the inverse of this duality mapping induced by the $\hat{\mathbf{n}} \times$ operator.
Indeed, such a inverse map is essential for the consistent discretization of products of integral operators. Although the higher-order case was not explored in the numerical experiments of \cite{christiansen2002}, the construction of the map $\Theta$ can be readily generalized to higher-order functions such as the GWP functions.

More recently, a similar definition of $\Theta$ have been presented in \cite{Bourhis2024higher} and \cite{Bouhris2025}, leveraging higher-order Helmholtz projectors in both the primal and auxiliary spaces. This map was used to enable higher-order discretizations of the EFIE, MFIE, and CFIE.

In \cite{Bernd2024ho}, b-spline higher-order dual functions are constructed for the conforming testing of the MFIE. 

A common feature of most of these approaches is the use of quasi-Helmholtz decompositions. Indeed, the separation into solenoidal and non-solenoidal subspaces is typically required to handle low-frequency breakdown and to obtain a discretization in which the square of the hypersingular part of the single-layer operator vanishes.

In this contribution, we use the $\Theta$ map defined in \cite{christiansen2002} in the discretization of the single-source equation. Preliminary results on the lowest-order discretization with the $\Theta$ map were presented in \cite{Zuccotti2024} and \cite{Zuccotti2025MFIE}. Here we leverage the flexibility in the choice of the primal and auxiliary spaces, to include higher-order GWP spaces. 

We present numerical experiments in which the proposed single-source ansatz is used for both the lowest-order RWG-BC discretization and higher-order discretizations involving GWP primal functions, and high-order dual functions that are implicitly defined as the image of the $\Theta$ operator. In all cases, we compare the conventional discretization, where one of the physical traces is expanded in the dual space, with the novel proposed formulation, in which both electric and magnetic traces are represented in the primal space.

Furthermore, we discuss a lowest-order discretization that leverages the $\Theta$ map to obtain a computationally cheaper discretization, using only RWG functions, as opposed to the discretization using RWG and BC functions, without jeopardising accuracy and efficiency.

We present an error analysis for the trace fields and for the near and far fields to demonstrate the accuracy and conditioning properties of the proposed methods.

The paper is organized as follows:
\begin{itemize}
\item Section~\ref{sec:continuous} recalls the continuous single-source equation, and Section~\ref{sec:discretization} reviews its conventional discretization.
\item Section~\ref{sec:new_ansatz} presents the novel ansatz that allows both traces to be expanded in the primal space.
\item Section~\ref{sec:Nedelec_projectors} introduces the $\Theta$ map and its properties.
\item Section~\ref{sec:novel} presents the discretization with $\Theta$, and Section~\ref{sec:numericalexp} reports numerical results for different choices of spaces and geometries.
\end{itemize}

\section{Single-Source Integral Equation}\label{sec:continuous}
We consider a homogeneous domain $\Omega_1\subset \mathbb{R}^3$ with permittivity $\epsilon_1=\epsilon_r\epsilon_0$ and permeability $\mu_1=\mu_r\mu_0$ immersed in a background medium $\Omega_0:=\mathbb{R}^3\setminus\bar{\Omega}_1 $, characterized by permittivity $\epsilon_0$ and permeability $\mu_0$. The structure is illuminated by an incident time-harmonic electromagnetic field $(\mathbf{e}^{\mathrm{inc}}, \mathbf{h}^{\mathrm{inc}})$ generated by an external source. We assume the incident field to be a plane wave
\begin{equation}
\begin{split}
\mathbf e^{\mathrm{inc}}(\mathbf x)
&=
\mathbf{\hat{e}} \, e^{-j\boldsymbol{\kappa}_0\cdot \mathbf x},
\\
\mathbf h^{\mathrm{inc}}(\mathbf x)
&=
\frac{1}{\eta_0}\,
\boldsymbol{\hat{\kappa}}_0\times \mathbf{\hat{e}} \,
e^{-j \boldsymbol\kappa_0\cdot \mathbf x},\quad \mathbf x \in \Omega_0,
\end{split}
\end{equation}
with $\mathbf{\hat{e}}$ the polarization direction, $\boldsymbol\kappa_0=\kappa_0\boldsymbol{\hat{\kappa}}_0$ the wavevector, $\kappa_0$ the wavenumber and $\eta_0=\sqrt{\frac{\mu_0}{\epsilon_0}} $ the characteristic impedance of the external background medium.
The time dependence $e^{j\omega t}$ is assumed and will be omitted in the rest of the work. Let 
$(\mathbf{e}\times\mathbf{\hat{n}}, \mathbf{\hat{n}}\times\mathbf{h})$ denote the electric and magnetic tangential traces on the boundary $\Gamma:=\partial\Omega_1$, where $\mathbf{\hat{n}}$ is the outward unit normal to $\Gamma$.
Since no physical currents are present in dielectric media, the tangential field components are continuous across the interface, i.e. $(\mathbf{e}^+\times\mathbf{\hat{n}},\mathbf{\hat{n}}\times\mathbf{h}^+)=(\mathbf{e}^-\times\mathbf{\hat{n}},\mathbf{\hat{n}}\times\mathbf{h}^-)=(\mathbf{e}\times\mathbf{\hat{n}},\mathbf{\hat{n}}\times\mathbf{h})$.
We recall the definition of the Maxwell single and double layer operators
\begin{equation}
    \begin{split}
        \mathcal{T}(\kappa)&=\mathcal{T}_s(\kappa)+\mathcal{T}_h(\kappa),\\
            (\mathcal{T}_s(\kappa)\mathbf{j})(\mathbf{r})&=-
        j\kappa\mathbf{\hat{n}}\times\int_\Gamma\frac{e^{-j\kappa R}}{4\pi R}\mathbf{j}(\mathbf{r'})dS',\\
        (\mathcal{T}_h(\kappa)\mathbf{j})(\mathbf{r})&=\frac{1}{j\kappa}\mathbf{\hat{n}}\times\mbox{p.v.}\int_\Gamma\nabla\frac{e^{-j\kappa R}}{4\pi R}\nabla' \cdot\mathbf{j}(\mathbf{r'})dS',\\
          (\mathcal{K}(\kappa)\mathbf{j})(\mathbf{r})&=\mathbf{\hat{n}}\times \mbox{p.v.}\int_\Gamma\nabla \times \frac{e^{-j\kappa R}}{4\pi R}\mathbf{j}(\mathbf{r'})dS',
    \end{split}
\end{equation}
with $R=|\mathbf{r}-\mathbf{r'}|$. In the absence of internal sources, the tangential traces associated with a regular Maxwell solution $(\mathbf{e},\mathbf{h})$ in $\Omega_1$ satisfy the interior Calderón formula and can be represented as
\begin{equation}\label{interior_repformula}
    \begin{split}
        \begin{pmatrix}
            -\mathcal{K}(\kappa_1)+\frac{1}{2} & \eta_1\mathcal{T}(\kappa_1) \\
            -\frac{1}{\eta_1}\mathcal{T}(\kappa_1) & -\mathcal{K}(\kappa_1)+\frac{1}{2}
        \end{pmatrix}
        \begin{pmatrix}
  \mathbf{p} \\
            \mathbf{q}
        \end{pmatrix}
        =
        \begin{pmatrix}
            \mathbf{e}\times \mathbf{\hat{n}}  \\
            \mathbf{\hat{n}} \times \mathbf{h}
        \end{pmatrix},
    \end{split}
\end{equation}
with $\kappa_1=\sqrt{\epsilon_{r} \mu_{r}}\kappa_0,\ \eta_1=\sqrt{\frac{\mu_{r}}{\epsilon_{r}}}\eta_0$ the wavenumber and the characteristic impedance of the internal medium.
The pair $(\mathbf{p},\mathbf{q})=(\mathbf{e}\times \mathbf{\hat{n}},\mathbf{\hat{n}}\times\mathbf{h})$ itself satisfies \eqref{interior_repformula}, but it is not a unique solution. 
Likewise, in the presence of an incident field, the extinction formula for the exterior traces reads
\begin{equation}\label{externalformula}
    \begin{split}
        \begin{pmatrix}
            \mathcal{K}(\kappa_0)-\frac{1}{2} & -\eta_0\mathcal{T}(\kappa_0) \\
            \frac{1}{\eta_0}\mathcal{T}(\kappa_0) & \mathcal{K}(\kappa_0)-\frac{1}{2}
        \end{pmatrix}
        \begin{pmatrix}
              \mathbf{e} \times \mathbf{\hat{n}}\\
            \mathbf{\hat{n}} \times \mathbf{h}
        \end{pmatrix}
        =
        \begin{pmatrix}
            \mathbf{\hat{n}} \times \mathbf{e}^{\mathrm{inc}} \\
            \mathbf{h}^{\mathrm{inc}}\times \mathbf{\hat{n}} 
        \end{pmatrix}.
    \end{split}
\end{equation}
Substituting \eqref{interior_repformula} into \eqref{externalformula} yields 
\begin{equation}\label{dualsource}
    \begin{split}
        &\begin{pmatrix}
            \mathcal{K}_0-\frac{1}{2} & -\eta_0\mathcal{T}_0 \\
            \frac{1}{\eta_0}\mathcal{T}_0 & \mathcal{K}_0-\frac{1}{2}
        \end{pmatrix}
        \begin{pmatrix}
            -\mathcal{K}_1+\frac{1}{2} & \eta_1\mathcal{T}_1 \\
            -\frac{1}{\eta_1}\mathcal{T}_1 & -\mathcal{K}_1+\frac{1}{2}
        \end{pmatrix}
        \begin{pmatrix}
              \mathbf{p} \\
            \mathbf{q}
        \end{pmatrix}=\\
        &=\begin{pmatrix}
            \mathbf{\hat{n}} \times \mathbf{e}^{\mathrm{inc}} \\
            \mathbf{h}^{\mathrm{inc}}\times \mathbf{\hat{n}} 
        \end{pmatrix}.
    \end{split}
\end{equation}
where we introduce the abbreviations $\mathcal{T}_i,\mathcal{K}_i=\mathcal{T}(\kappa_i),\mathcal{K}(\kappa_i)$.
To reduce the system to a single-source equation, we exploit the redundancy in $(\mathbf{p}, \mathbf{q})$ by selecting either a magnetic ansatz $(\mathbf{p}, \mathbf{0})$ or an electric ansatz $(\mathbf{0}, \mathbf{q})$.
Here, we adopt the magnetic ansatz and we select the first row of the resulting system, which yields
\begin{equation}\label{singlesource}
    \begin{split}
           \bigg(\Big(\mathcal{K}_0-\frac{1}{2}\Big)\Big(-\mathcal{K}_1+\frac{1}{2}\Big) +\frac{\eta_0}{\eta_1}\mathcal{T}_0\mathcal{T}_1\bigg)\  \mathbf{p}\ =\mathbf{\hat{n}} \times \mathbf{e}^{\mathrm{inc}}.
    \end{split}
\end{equation}
The operator in the LHS of \eqref{singlesource} exhibits a well-conditioned spectrum bounded away from both zero and infinity. Indeed, both the two terms in \eqref{singlesource} are well conditioned as it is discussed in \cite{Valdes2011}.

Under this ansatz, $\mathbf{p}$ represents an auxiliary surface current, while the physical traces are recovered through the internal operators:
\begin{equation}
\mathbf{\hat{n}} \times \mathbf{h}=-\frac{1}{\eta_1}\mathcal{T}_1 \ \mathbf{p},\quad \mathbf{e}\times\mathbf{\hat{n}} =\Big(-\mathcal{K}_1+\frac{1}{2}\Big)\mathbf{p}.
\end{equation}
\section{Discretization with Primal and Dual Functions}\label{sec:discretization}
We recall here the discretization of \eqref{singlesource}.
Following the Petrov–Galerkin method, each integral operator is discretized with respect to a set of basis and test functions. We adopt the following notation: for each integral operator $\mathcal{O}$, the matrix elements of the discretized operator are defined by
$(\mathcal{O}_{xy})_{ij}=\langle \mathbf{\hat{n}} \times \mathbf{x}_i, \mathcal{O}\mathbf{y}_j\rangle$,
with $\mathbf{x}_i$ and $\mathbf{y}_i$ the $i^{\mathrm{th}}$ basis functions of two div-conforming spaces, and $\langle\cdot,\cdot\rangle$ is the $L^2$ inner product.
Primal and dual spaces are required for the conforming discretization of the single- and double-layer operators.
The most common choice is the RWG and the BC spaces as primal and dual spaces, respectively. Here we denote by $\mathbf{f}_i$ and $\mathbf{g}_i$ the $i^{\mathrm{th}}$ RWG and BC basis functions. Consequently, the indices $f$ and $g$ indicate expansions in the RWG and BC spaces, respectively.  

We denote by $G_{gf}$ the map $(G_{gf})_{ij}=\langle \mathbf{\hat{n}} \times \mathbf{g}_i, \mathbf{f}_j\rangle$, i.e., the mixed Gram matrix. Its inverse is denoted by $G^{fg}$. $G_{gf}$ is the discrete duality pairing between curl-conforming and div-conforming spaces. Its inverse $G^{fg}$ or the adjoint of the inverse ${G^{fg}}^\dagger$ is required in the discretization of products of Maxwell integral operators, as such operators are typically applied to div-conforming functions and tested with curl-conforming functions. The discretization of the product of two integral operators is then given by $G^{fg}\mathcal{O}_{gg}({-G^{fg}}^\dagger)\mathcal{O}_{ff}$ or $G^{fg}\mathcal{O}_{gf}{G^{fg}}\mathcal{O}_{gf}$.

If the auxiliary current $\mathbf{p}$ is expanded with RWG functions, the conforming discretization of \eqref{singlesource} reads
\begin{equation}\label{discrete_signlesoure}
\begin{split}
&G^{fg}\bigg(\Big(\mathcal{K}_0-\frac{1}{2}\Big)_{gf}G^{fg} \Big(-\mathcal{K}_1+\frac{1}{2}\Big)_{gf}+\\
&+\frac{\eta_0}{\eta_1}(\mathcal{T}_0)_{gg}(-{G^{fg}}^\dagger)(\mathcal{T}_1)_{ff}\bigg)\ \mathbf{p}^f
=G^{fg}(\mathbf{\hat{n}} \times \mathbf{e}^{\mathrm{inc}})_g,
\end{split}
\end{equation}
where $\mathbf{p}^f$ is the unknown vector of coefficients that expands $\mathbf{p}$ in the RWG space, and $((\mathbf{\hat{n}} \times \mathbf{e}^{\mathrm{inc}})_g)_i=\langle \mathbf{\hat{n}} \times \mathbf{g}_i,\mathbf{\hat{n}} \times \mathbf{e}^{\mathrm{inc}}\rangle$.

This conforming discretization ensures a regular and well-conditioned matrix system and accurate results for the unknown $\mathbf{p}$. However, we note that the physical traces are computed by applying the first pair of internal operators to the solution of \eqref{discrete_signlesoure}:
\begin{equation}
\label{physical_bc}
\begin{split}
(\mathbf{\hat{n}} \times \mathbf{h})^g&=\frac{1}{\eta_1}{G^{fg}}^
\dagger(\mathcal{T}_1)_{ff}\mathbf{p}^f,\\
(\mathbf{e}\times\mathbf{\hat{n}})^f&=G^{fg}\big(-\mathcal{K}_1+\tfrac{1}{2}\big)_{gf}\mathbf{p}^f.
    \end{split}
\end{equation}
\begin{figure*}[!t]
    \centering
    \begin{subfigure}{0.5\textwidth}
        \centering
        \includegraphics[width=\linewidth]{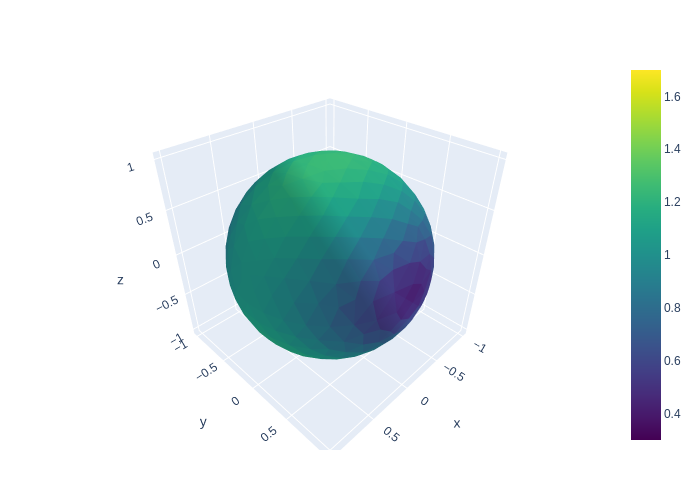}
        \caption{Magnitude of the magnetic trace (A/m) expanded in RWG functions.}
        
    \end{subfigure}\hfill
    \begin{subfigure}{0.5\textwidth}
        \centering
        \includegraphics[width=\linewidth]{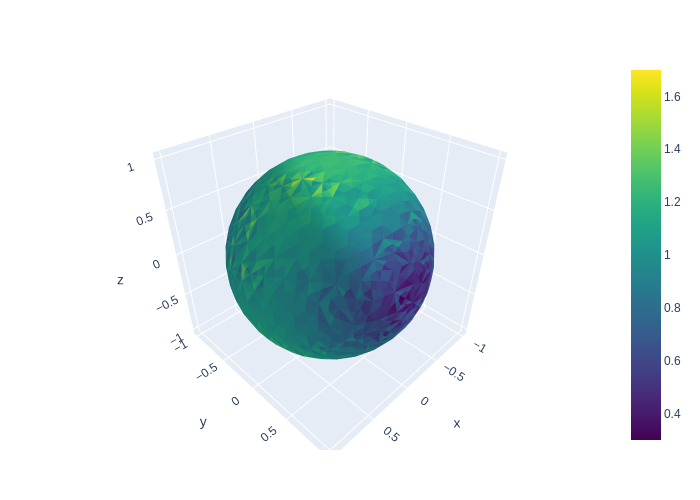}
        \caption{Magnitude of the magnetic trace (A/m) expanded in BC functions.}
    \end{subfigure}
    \caption{Comparison of the magnitude of the magnetic trace for the transmission problem expanded in RWGs (left) and in BCs (right) for a sphere. The representation in RWGs is obtained by solving the PMCHWT equation and the representation in BCs by solving the single-source equation. The inferior approximation accuracy of the BC space is already visible for this smooth surface example.}
    \label{fig:mag_traces_3d_RWG_BC_sphere}
\end{figure*}
\begin{figure*}[!t]
    \centering
    \begin{subfigure}{0.5\textwidth}
        \centering
        \includegraphics[width=\linewidth]{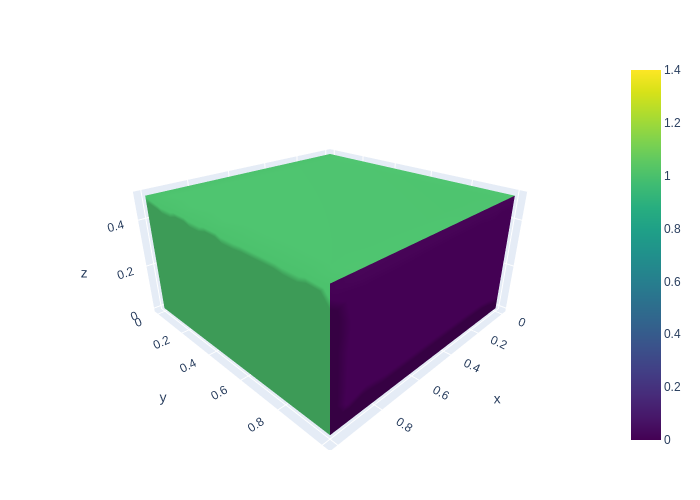}
        \caption{Magnitude of the magnetic trace (A/m) expanded in RWG functions.}
        
    \end{subfigure}\hfill
    \begin{subfigure}{0.5\textwidth}
        \centering
        \includegraphics[width=\linewidth]{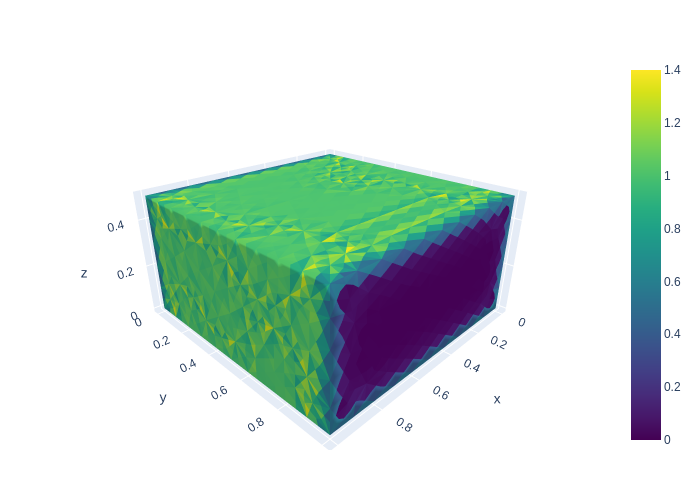}
        \caption{Magnitude of the magnetic trace (A/m) expanded in BC functions.}
    \end{subfigure}
    \caption{Comparison of the magnitude of the magnetic trace for the transmission problem expanded in RWGs (left) and in BCs (right) for a cuboid with $\eta_1,\kappa_1=\eta_0,\kappa_0$. The representation in the RWGs is obtained by projecting the incident magnetic trace $\mathbf{\hat{n}} \times \mathbf{h}^\mathrm{inc}$ onto the RWG space. The representation in the BC space is obtained by solving the single-source equation. The solution degradation in the BC space and the spurious continuity across the edges are evident for this example.}
    \label{fig:mag_traces_3d_RWG_BC_cuboid_air}
\end{figure*}
This implies that one of the two traces is expanded in BC functions defined on the barycentric refinement. This situation does not occur in other common BIEs such as the EFIE or the MFIE, where BC functions are used to precondition or to test such equations, whereas the physical unknown remains expanded in terms of RWG functions.

As we mentioned in the introduction, approximation errors in the $H^{0}(\operatorname{div}; \Gamma)$ norm of quantities expanded in BC functions can be larger than those obtained with RWG functions. In addition, BC basis functions model currents that flow along an edge of the mesh (as opposed to across an edge) and as a result enforce continuity across element interfaces. In the presence of sharp geometric features, this imposed continuity may be nonphysical, since electric and magnetic traces may exhibit discontinuities that occur along abrupt variations of the surface normal.

To demonstrate this effect, we show two comparisons of the same quantity for a sphere and a cuboid geometry expanded either in the primal RWG space or in the dual BC space. In Fig.~\ref{fig:mag_traces_3d_RWG_BC_sphere} we show the magnitude of the magnetic trace obtained by solving \eqref{discrete_signlesoure} and computing \eqref{physical_bc} with $(\eta_1,\kappa_1)=(\frac{\eta_0}{1.55},1.55\kappa_0)$, first expanded in terms of RWG functions (on the left), and then the same trace expanded in terms of BC functions (on the right), obtained by solving the PMCHWT equation.

In Fig.~\ref{fig:mag_traces_3d_RWG_BC_cuboid_air} we show magnitude of the magnetic traces obtained for a cuboid occupied by the same material as the background ($\eta_1,\kappa_1={\eta_0},\kappa_0$). In this special case, the exact solution is known. It corresponds to the trace of the incident field, which has piecewise constant magnitude on the faces of the cuboid.

The spurious continuity across the edges is evident in Fig.~\ref{fig:mag_traces_3d_RWG_BC_cuboid_air}. In both the sphere and the cuboid cases, the expansion in the dual space visibly affects the representation of the approximated quantity. In Section \ref{sec:numericalexp} we will show a detailed analysis of the accuracy obtained for both these geometries.
In the next section, we discuss possible strategies for obtaining both physical traces expanded in the RWG space.

\section{Novel Ansatz for the Single-Source Equation\label{sec:new_ansatz}}

To get both traces expanded in the primal space, one might consider alternative discretization schemes for the integral operators in \eqref{discrete_signlesoure}. 
In particular, one possibility is to choose an asymmetric mixed discretization of the single-layer operators such as the one in \cite{Yan_muller_BC} for the M\"uller equation. However, the discretization of the single-layer operator with RWG basis functions and rotated BC test functions is less accurate than the one in \eqref{discrete_signlesoure}. The M\"uller equation is a notable exception because the specific combination of the two single-layer operators weakens their singularity, thereby mitigating the error introduced by the mixed discretization.

Similarly, nonconforming discretizations of the double-layer and identity operators could lead to a system in which both traces are expanded in the primal space. However, this would introduce additional errors, since the mixed conforming scheme has been shown to be the most accurate \cite{CoolsMFIEconforming}. The PMCHWT equation is an exception, where a symmetric discretization of the double-layer operator is admissible since the identity operator cancels in the combined representation formulas.

Instead, the single-source equation offers unique flexibility in the choice of the ansatz. From \eqref{dualsource}, one can select either the ansatz $(\mathbf{p},\mathbf{0})$ for the auxiliary unknown, leading to \eqref{discrete_signlesoure}, or the alternative $(\mathbf{0},\mathbf{q})$. From these two ans\"atze, two single-source equations can be derived. In particular, by selecting the first row of \eqref{dualsource} in the ansatz $(\mathbf{p},\mathbf{0})$ and the second row of \eqref{dualsource} in the ansatz $(\mathbf{0},\mathbf q)$, the following single-source equations are obtained:
\begin{equation}\label{double_discrete_singlesource}
\begin{split}
&G^{fg}\bigg(\Big(\mathcal{K}_0-\frac{1}{2}\Big)_{gf}G^{fg} \Big(-\mathcal{K}_1+\frac{1}{2}\Big)_{gf}+\\
&+\frac{\eta_0}{\eta_1}(\mathcal{T}_0)_{gg}(-{G^{fg}}^\dagger)(\mathcal{T}_1)_{ff}\bigg)\ \mathbf{p}^f
=G^{fg}(\mathbf{\hat{n}} \times \mathbf{e}^{\mathrm{inc}})_g,\\
&G^{fg}\bigg(\Big(\mathcal{K}_0 - \frac{1}{2}\Big)_{gf} G^{fg} \Big(-\mathcal{K}_1 + \frac{1}{2}\Big)_{gf} + \\
&+ \frac{\eta_1}{\eta_0} (\mathcal{T}_0)_{gg} \big(-{G^{fg}}^\dagger\big) (\mathcal{T}_1)_{ff} \bigg) \mathbf{q}^f
= G^{fg} (\mathbf{h}^{\mathrm{inc}} \times \mathbf{\hat{n}})_g.
\end{split}
\end{equation}
The two physical traces can then be recovered in the primal space as
\begin{equation}
\label{traces_magnetic_double_bc}
\begin{split}
(\mathbf{\hat{n}} \times \mathbf{h})^f &= G^{fg}\big(-\mathcal{K}_1 + \tfrac{1}{2}\big)_{gf}\mathbf{q}^f,\\
(\mathbf{e}\times\mathbf{\hat{n}})^f &= G^{fg}\big(-\mathcal{K}_1 + \tfrac{1}{2}\big)_{gf}\mathbf{p}^f.
\end{split}
\end{equation}
Although the two equations in \eqref{double_discrete_singlesource} are distinct, they share the same discretized integral operators. Hence, solving both systems is a viable strategy, since the discretization of the integral operators is typically the most expensive step. The additional computational cost of this double-ansatz formulation is essentially equivalent to doubling the number of iterations. The conforming discretizations \eqref{double_discrete_singlesource} of the single-source integral equation are known to result in linear systems whose solution by Krylov methods only requires a small number of Krylov iterations. 

In the remainder of this paper, additional strategies are discussed to further optimize the efficiency of the solver.
In the following sections, we introduce an alternative discretization strategy designed to preserve favorable conditioning properties while enabling a more flexible choice of the primal and dual space, beyond the conventional RWG-BC spaces.

\section{General Dual-Space Construction through Quasi-Helmholtz Decomposition}\label{sec:Nedelec_projectors}

As mentioned in the previous section, the BC space is required to obtain an invertible, well-conditioned map for the mixed Gram matrix $(G_{gf})_{ij} = \langle \mathbf{\hat{n}}\times \mathbf{g}_i, \mathbf{f}_j \rangle$. Moreover, the BC space is designed such that $P_{\Sigma,f} G^{fg} P_{\Sigma,g}^\dagger = 0$ \cite{BUFFA2005461}, where $P_{\Sigma,f}$ and $P_{\Sigma,g}$ denote the quasi-Helmholtz projectors in the respective star subspaces of the RWG and BC spaces (we will give a definition of the star projector later in this section).
From this property it follows that $G^{fg}(\mathcal{T}_h)_{gg}{G^{fg}}^\dagger(\mathcal{T}_h)_{ff} = 0$, with $\mathcal{T}_h$ denoting the hypersingular part of the single-layer operator.  

From now on, we will use the notation with indices $f$ and $g$ not only for the RWG and BC spaces but also to denote two general div-conforming spaces. We will refer to these as the primal and the auxiliary spaces, respectively.
Let $\mathbf{f}_i$ denote the set of primal div-conforming basis functions, and $\mathbf{g}_i$ the set of auxiliary div-conforming basis functions.
We denote by $A_{ff}$ and $A_{gg}$ the Gram matrices $(A_{ff})_{ij} = \langle \mathbf{f}_i, \mathbf{f}_j \rangle$, $(A_{gg})_{ij} = \langle \mathbf{g}_i, \mathbf{g}_j \rangle$, and by $A^{ff}$ and $A^{gg}$ their inverses. Note the absence of the $\mathbf{\hat{n}} \times$ factor in these pairings, in contrast to the notation we used for the discretized integral operators. 

For a general choice of the spaces, $G_{gf}$ is not invertible. The map $A^{ff}(-G_{gf}^{\dagger}) A^{gg}$ may be used in place of the inverse Gram matrix, but  $P_{\Sigma,f} A^{ff}(-G_{gf}^{\dagger}) A^{gg} P_{\Sigma,g}^\dagger \ne 0$ for general div-conforming spaces. 

In the following, we want to define a map $\Theta_{gf}$, and its dual counterpart $\Theta^{fg} = A^{ff}(-\Theta_{gf}^\dagger) A^{gg}$, that substitutes the usual $G_{gf}$ and $G^{fg}$, but that allows a general choice of the spaces and satisfies $P_{\Sigma,f}\Theta^{fg} P_{\Sigma,g}^\dagger = 0$ as well.

In \cite{christiansen2002}, such a map $\Theta_{gf}$ is obtained by leveraging the quasi-Helmholtz decomposition.
In this section, we review the implementation of $\Theta_{gf}$ following the definition given in \cite{christiansen2002}, and consistent with our notation. We note that the operator denoted by $\Theta_h$ in \cite{christiansen2002} corresponds to the transpose of the map that we will denote as $\Theta^{fg}$.    

In order to define $\Theta_{gf}$ and $\Theta^{fg}$, we will need the following:
\begin{itemize}
\item {Let $\mathcal{F}_{h}$ be the primal space defined on a mesh of size $h$, and $\mathbf{f}_i$ the $i^{\mathrm{th}}$ basis function of $\mathcal{F}_h$. $\mathcal{F}_h$ is typically the RWG space or the GWP($p$) space of order $p \ge 0$ in higher-order discretizations.
We also introduce the smallest Lagrange scalar discontinuous space $\mathcal{C}_{\mathcal{F},h}$ that contains the image of the divergence operator acting on $\mathcal{F}_{h}$. If $\mathcal{F}_{h}$ is the GWP($p$) space, $\mathcal{C}_{\mathcal{F},h}$ is the space of scalar discontinuous Lagrange functions of the same order.    
Then the star matrix $\Sigma_{fc}$ is defined by $({\Sigma_{fc}})_{ij} = \langle \mathbf{\nabla}\cdot\mathbf{f}_i, c_j \rangle$, where $c_j$ is the $j^{\mathrm{th}}$ basis function of $\mathcal{C}_{\mathcal{F},h}$ on the same mesh.
We define the quasi-Helmholtz star projector as
\begin{equation}\label{star_proj_def}
P_{\Sigma,f}=A^{ff}\Sigma_{fc}(\Sigma_{fc}^\dagger A^{ff}\Sigma_{fc})^+\Sigma_{fc}^\dagger,
\end{equation}
with $^+$ denoting the pseudoinverse. $P_{\Sigma,f}$ projects onto the $L^2$ orthogonal complement of the subspace of divergence free functions in the primal space.}

\item {We introduce the auxiliary space $\mathcal{G}_h$ that will have the analogous role of the Buffa–Christiansen space.  
This auxiliary space can be chosen quite generally, provided it is div-conforming. 
We will see in the numerical experiments that the auxiliary space can coincide with $\mathcal{F}_h$ itself 
or it can be chosen as a higher-order GWP space, including semi-integer order spaces as the BDM family \cite{BDM1985}, or as a space of the same order as $\mathcal{F}_h$ but defined on a refined mesh.  
We also introduce the largest scalar continuous Lagrange space $\mathcal{P}_{\mathcal{G},h}$ such that the image of the surface-curl operator $\mathbf{\hat{n}} \times \mathbf{\nabla}$ acting on $\mathcal{P}_{\mathcal{G},h}$ is in $\mathcal{G}_h$. If $\mathcal{G}_h$ is the GWP($p$) space, $\mathcal{P}_{\mathcal{G},h}$ is the continuous Lagrange space of order $p+1$. Then we define the loop matrix $\Lambda_{gp}$
\begin{equation}
    \begin{split}
        &(\Lambda_{gp})_{ij} = \langle \mathbf{g}_i, \mathbf{\hat{n}}\times\mathbf{\nabla} p_j \rangle=-\langle \mathbf{\hat{n}}\times\mathbf{g}_i, \mathbf{\nabla} p_j \rangle,\\
        &\ \mathbf{g}_i \in \mathcal{G}_h,\ p_j \in \mathcal{P}_{\mathcal{G},h}.
    \end{split}
\end{equation}}
\item{We also define the $L^2$ projection from the discontinuous primal scalar space onto the continuous auxiliary scalar space previously introduced. This projection is performed by the matrix 
\begin{equation}\label{proj_cont}
    P^{p}_{\ c} = A^{pp}A_{pc},
\end{equation}
with
\begin{equation}
    \begin{split}
    (A_{pc})_{ij} &= \langle p_i, c_j \rangle,\\
    A^{pp} &= A_{pp}^{-1},\\
    (A_{pp})_{ij} &= \langle p_i, p_j \rangle, \quad p_i \in \mathcal{P}_{\mathcal{G},h},\ c_j \in \mathcal{C}_{\mathcal{F},h}.
    \end{split}
\end{equation}}
\end{itemize}
In order to build the map $\Theta_{gf}$, we consider the usual mixed Gram matrix $(G_{gf})_{ij} = \langle \mathbf{\hat{n}}\times \mathbf{g}_i, \mathbf{f}_j \rangle$ and we decompose it in two terms, corresponding to the Helmholtz components of the primal space:
\begin{equation}\label{mixed_helmholtz_components}
    \begin{split}
        G_{gf} &= G_{gf}(\mathcal{I}^f_{\ f}-P_{\Sigma,f}) + G_{gf}P_{\Sigma,f},
    \end{split}
\end{equation}
with $\mathcal{I}^f_{\ f}$ denoting the identity matrix.
The map $\Theta_{gf}$ is constructed by modifying the second term in \eqref{mixed_helmholtz_components}. 
The term $G_{gf}P_{\Sigma,f}$ is replaced by
\begin{equation}\label{substitution_star_comp}
    G_{gf}P_{\Sigma,f} \to \Lambda_{gp} P^{p}_{\ c}(\Sigma_{fc}^\dagger A^{ff}\Sigma_{fc})^+\Sigma_{fc}^\dagger.
\end{equation}
Thus, we define
\begin{equation}\begin{split}
\label{theta_projectors}
    \Theta_{gf}
        &= G_{gf}(\mathcal{I}^f_{\ f}-P_{\Sigma,f}) + \Lambda_{gp} P^{p}_{\ c}(\Sigma_{fc}^\dagger A^{ff}\Sigma_{fc})^+\Sigma_{fc}^\dagger,
\end{split}
\end{equation}
and its dual
\begin{equation}
 \Theta^{fg} = A^{ff}(-{\Theta_{gf}}^\dagger) A^{gg}.
\end{equation}

Given the definition of $P_{\Sigma,f}$, 
the substitution in \eqref{substitution_star_comp} is equivalent to replace the chain  $G_{gf}A^{ff}\Sigma_{fc}$ with $\Lambda_{gp} P^{p}_{\ c}$. In a discrete sense, $G_{gf}A^{ff}\Sigma_{fc}$ represents the composition of the weak gradient operator with the $\mathbf{\hat{n}}\times$, whereas $\Lambda_{gp} P^{p}_{\ c}$ represents the composition of the $L^2$ projection onto the continuous scalar space with the exact $\mathbf{\hat{n}}\times \nabla$ operator. 

As desired, the map $\Theta^{fg}$ satisfies
\begin{equation}
\begin{split}
    &P_{\Sigma,f}\Theta^{fg}P_{\Sigma,g}^\dagger =\\
    &=P_{\Sigma,f}A^{ff}(\Sigma_{fc}(\Sigma_{fc}^\dagger A^{ff}\Sigma_{fc})^+{P^{p}_{\ c}}^\dagger\Lambda_{gp}^\dagger)A^{gg}P_{\Sigma,g}^\dagger = 0,
\end{split}
\end{equation}
since $P_{\Sigma,f} A^{ff} (\mathcal{I}^f_{\ f}-P_{\Sigma,f})^\dagger = 0$ and  $\Lambda_{gp}^\dagger A^{gg}P_{\Sigma,g}^\dagger  = 0$.

It is worth noting that in the original formulation in \cite{christiansen2002}, the quasi-Helmholtz decomposition was performed through the inversion of a saddle-point system. Indeed, the following relation holds:
\begin{equation}\label{schur_complement_expression}
\begin{split}
    &\begin{pmatrix}
        A_{ff} & \Sigma_{0,fc}\\
        \Sigma_{0,fc}^\dagger & 0
    \end{pmatrix}^{-1}
    \begin{pmatrix}
			A^{ff}\\
			0
			\end{pmatrix} =\\
    &=\begin{pmatrix}
       A^{ff}(A_{ff}-\Sigma_{fc}(\Sigma_{fc}^\dagger A^{ff}\Sigma_{fc})^+\Sigma_{fc}^\dagger)\\
       (\Sigma_{0,fc}^\dagger A^{ff}\Sigma_{0,fc})^{-1}\Sigma_{0,fc}^\dagger
   \end{pmatrix}
   =\\
   &=\begin{pmatrix}
       (\mathcal{I}^f_{\ f}-P_{\Sigma,f})\\
       (\Sigma_{0,fc}^\dagger A^{ff}\Sigma_{0,fc})^{-1}\Sigma_{0,fc}^\dagger
   \end{pmatrix},
\end{split}
\end{equation}
where $\Sigma_{0,fc}$ is defined as $(\Sigma_{0,fc})_{ij} = \langle \nabla \cdot \mathbf{f}_i, c_j \rangle$, $c_j \in \mathcal{C}_{0,\mathcal{F},h}$, and $\mathcal{C}_{0,\mathcal{F},h}$ is the space of piecewise constant Lagrange functions with zero global mean. In the LHS of \eqref{schur_complement_expression} we show the saddle-point system reported in \cite{christiansen2002}, whereas the RHS consists of the same projectors we used in the definition of $\Theta_{gf}$.
Hence, we can write $\Theta_{gf}$ also in the form
\begin{equation}\label{theta_nedelec_saddle}
\Theta_{gf}=\begin{pmatrix}
G_{gf} &  \Lambda_{gp} P^{p}_{\ c}
\end{pmatrix}
\begin{pmatrix}
        A_{ff} & \Sigma_{0,fc}\\
        \Sigma_{0,fc}^\dagger & 0
    \end{pmatrix}^{-1}
    \!\begin{pmatrix}
			A_{ff}\\
			0
			\end{pmatrix}.
\end{equation}
Although the definitions in \eqref{theta_projectors} and \eqref{theta_nedelec_saddle} are analytically equivalent, their numerical implementation has different key aspects. 

Lastly, we mention that recently, a similar definition of $\Theta^{fg}$ was presented in \cite{Bourhis2024higher} and \cite{Bouhris2025}, in the context of higher-order EFIE, MFIE, and CFIE. 
We denote with $\tilde\Theta^{fg}$ such a map. In our notation, $\tilde\Theta^{fg}$ is defined by
\begin{equation}\label{Bourhis_theta}
    \begin{split}
        \tilde{\Theta}^{fg}&=A^{ff}(-{\tilde{\Theta}_{gf}}^\dagger)A^{gg}\\
        \tilde{\Theta}_{gf}&=G_{gf}-P_{\Sigma,g}^\dagger G_{gf}P_{\Sigma,f}.
    \end{split}
\end{equation}
Hence the desired condition $P_{\Sigma,f} \tilde{\Theta}^{fg}P_{\Sigma,g}^\dagger=0$ can be readily verified. In this case, the implementation of the star quasi-Helmholtz projector is required in both the primal and the auxiliary space. The implementation of the quasi-Helmholtz projectors require either the inversion of the saddle-point system as in \eqref{schur_complement_expression} or the computation of the pseudo-inverse in \eqref{star_proj_def}, hence it is an expensive operation. An iterative inversion is possible but it requires a preconditioning strategy in both implementations. The definition of $\Theta_{gf}$ can be computationally preferable with respect to $\tilde{\Theta}_{gf}$ as it requires only the implementation of $P_{\Sigma,f}$ and does not require $P_{\Sigma,g}$ in this formulation. Instead, it requires the implementation of $\Lambda_{gp}$, which is a sparse matrix, and the inversion of $A_{pp}$ which is sparse symmetric positive definite and well-conditioned.

\subsection{Efficient implementation of the Helmholtz projectors}
\label{sec:Efficient implementation of the Helmholtz projectors}

For lowest-order discretizations, algebraic multigrid techniques can be employed to efficiently approximate the action of the operator $(\Sigma_{fc}^\dagger A^{ff}\Sigma_{fc})^+$ \cite{livne_lean_2012, napov_algebraic_2012}. In higher-order settings, however, the corresponding operators lose sparsity and standard black-box multigrid strategies are no longer directly applicable.

We emphasize that efficient implementations of the Helmholtz projectors, including strategies tailored to higher-order discretizations, can be constructed. These rely on appropriate preconditioning of the associated saddle-point formulation on the LHS of \eqref{schur_complement_expression}.

A detailed analysis of these implementation aspects is beyond the scope of the present work and will be reported in another contribution. In this paper, we focus exclusively on the analytical properties and the role of the projectors within the proposed formulation.

\section{General Conforming Discretization of the Single-Source Equation with $\Theta$}\label{sec:novel}
We introduce here the discretization of the single-source equation that uses $\Theta^{fg}$ in place of the inverse of the mixed Gram matrix. 
Following the structure of \eqref{discrete_signlesoure}, under the single ansatz $(\mathbf{p},\mathbf{0})$ for the auxiliary unknown, $\Theta^{fg}$ is incorporated in the single-source equation as follows:  
\begin{equation}\label{single_source_theta_single_ansatz}
\begin{split}
    &{\Theta^{fg}}
    \bigg(
    \Big(\mathcal{K}_0-\frac{1}{2}\Big)_{g f}
    {\Theta^{fg}}
    \Big(-\mathcal{K}_1+\frac{1}{2}\Big)_{g f}
    +\\
    &+\frac{\eta_0}{\eta_1}
    (\mathcal{T}_0)_{g g}
    (-{\Theta^{f g}}^\dagger)
    (\mathcal{T}_1)_{f f}
    \bigg)
    \mathbf{p}^f
    =
    {\Theta^{f g}}
    (\mathbf{\hat{n}} \times \mathbf{e}^{\mathrm{inc}})_g.
    \end{split}
\end{equation}
Then the two physical traces are recovered as 
\begin{equation}\label{traces_theta_single_ansatz}
    \begin{split}
        (\mathbf{\hat{n}} \times \mathbf{h})^g&=\frac{1}{\eta_1}{\Theta^{fg}}^\dagger(\mathcal{T}_1)_{ff}\mathbf{p}^f,\\
        (\mathbf{e}\times\mathbf{\hat{n}})^f&=\Theta^{fg}\Big(-\mathcal{K}_1+\frac{1}{2}\Big)_{gf}\mathbf{p}^f,
    \end{split}
\end{equation}

In the scheme above, we preserve the structure of the conforming discretization, where the single-layer operators are discretized symmetrically, one in the primal space and the other in the auxiliary space, while the double-layer and the identity operators are both discretized using a mixed scheme. 
In short, the discretization in \eqref{single_source_theta_single_ansatz} is the same as in \eqref{discrete_signlesoure}, but with the BC space replaced by a general auxiliary space, and $G^{fg}$ replaced by $\Theta^{fg}$. The dual space is implicitly constructed in the range of ${\Theta^{fg}}^\dagger$, as a subspace of the auxiliary space. In contrast to the EFIE and the MFIE cases, in the single-source discretization we use the images of $\Theta^{fg}$ and ${\Theta^{fg}}^\dagger$ to expand the unknowns, rather than only to construct a preconditioner.   
 
In \eqref{traces_theta_single_ansatz}, the electric trace is expanded in the image of $\Theta^{fg}$, which lies in the primal space, whereas the magnetic trace is expanded in the image of ${\Theta^{fg}}^\dagger$, which lies in the auxiliary space. The approximation properties established in \cite{christiansen2002} only apply to expansions in the image of ${\Theta^{fg}}$. Unfortunately no similar approximation estimates have been proven for the range of ${\Theta^{fg}}^\dagger$. Hence, the two traces may exhibit different accuracy. The impact of this scheme on the accuracy is investigated through numerical experiments in Section \ref{sec:numericalexp}. 
A double ansatz $(\mathbf{p}, \mathbf{0})$-$(\mathbf{0},\mathbf{q})$ can be used again to get a discretization that allows both traces to be expanded in the image of $\Theta^{fg}$. 
In this case we get the two single-source systems
\begin{equation}\label{single_source_theta}
\begin{split}
    &{\Theta^{fg}}
    \bigg(
    \Big(\mathcal{K}_0-\frac{1}{2}\Big)_{g f}
    {\Theta^{fg}}
    \Big(-\mathcal{K}_1+\frac{1}{2}\Big)_{g f}
    +\\
    &+\frac{\eta_0}{\eta_1}
    (\mathcal{T}_0)_{g g}
    (-{\Theta^{f g}}^\dagger)
    (\mathcal{T}_1)_{f f}
    \bigg)
    \mathbf{p}^f
    =
    {\Theta^{f g}}
    (\mathbf{\hat{n}} \times \mathbf{e}^{\mathrm{inc}})_g.\\
    &{\Theta^{fg}}
    \bigg(
    \Big(\mathcal{K}_0-\frac{1}{2}\Big)_{g f}
    {\Theta^{fg}}
    \Big(-\mathcal{K}_1+\frac{1}{2}\Big)_{g f}
    +\\
    &+\frac{\eta_1}{\eta_0}
    (\mathcal{T}_0)_{g g}
    (-{\Theta^{f g}}^\dagger)
    (\mathcal{T}_1)_{f f}
    \bigg)
    \mathbf{q}^f
    =
    {\Theta^{f g}}
    (\mathbf{h}^{\mathrm{inc}} \times \mathbf{\hat{n}})_g,\\
\end{split}
\end{equation}
which share the same discretized integral operators. 

The two physical traces are then computed by
\begin{equation}\label{traces_theta}
    \begin{split}
        (\mathbf{\hat{n}} \times \mathbf{h})^f&=\Theta^{fg}\Big(-\mathcal{K}_1+\frac{1}{2}\Big)_{gf}\mathbf{q}^f,\\
        (\mathbf{e}\times\mathbf{\hat{n}})^f&=\Theta^{fg}\Big(-\mathcal{K}_1+\frac{1}{2}\Big)_{gf}\mathbf{p}^f.
    \end{split}
\end{equation}

In \eqref{single_source_theta_single_ansatz} and \eqref{single_source_theta} the discretized identity in the $\frac{1}{2}$ terms is represented by ${\Theta^{fg}}G_{gf}$, which does not coincide with the identity matrix. Spectral analysis of ${\Theta^{fg}}G_{gf}$ reveals that its largest eigenvalues remain close to unity, whereas the smallest eigenvalues are substantially smaller depending on the chosen auxiliary space (see Table \ref{tab:eigen_theta_gram}). We will see that this spectral property will impact the conditioning of the resulting system. In particular, the numerical results show that choosing the auxiliary space two orders higher than the primal space yields a well-conditioned system, as observed in \cite{Bourhis2024higher} and \cite{Bouhris2025} for the discretization based on $\tilde{\Theta}^{fg}$.  

\begin{table}[htbp]
\centering
\caption{Eigenvalues of $\Theta^{fg}G_{gf}$ for different primal and auxiliary spaces. The RWG-RWG case is characterized by a Kernel with $N_{\Sigma,f}-N_{\Lambda,f}$ vanishing eigenvalues, with $N_{\Sigma,f},\ N_{\Lambda,f}$ the dimension of the star and local loop subspaces of the RWG space.}
\label{tab:eigen_theta_gram}
\[
\begin{tabular}{|c|c|c|c|}
\hline
{RWG-RWG} & {RWG-BDM} & {RWG-GWP(1)} & {RWG-GWP(2)} \\
\hline
$\begin{array}{c}
1.0004\\
0.9999\\
\vdots \\
0.065\\
1.25 \times10^{-15}\\
\vdots \\
7.91 \times10^{-20}
\end{array}$
&
$\begin{array}{c}
0.995 \\
0.994 \\
0.994 \\
\vdots \\
\vdots \\
0.068 \\
0.063
\end{array}$
&
$\begin{array}{c}
0.996 \\
0.995 \\
0.993 \\
\vdots \\
\vdots \\
0.127 \\
0.120
\end{array}$
&
$\begin{array}{c}
0.998 \\
0.998 \\
0.998\\
\vdots \\
\vdots \\
0.706\\
0.705
\end{array}$\\
\hline

\end{tabular}
\]
\end{table}

\subsection{Ad hoc Scheme with the Identity Matrix.}
Lastly, we introduce an ad hoc scheme in which the single-source equation is discretized by using $\Theta^{fg}$ as in \eqref{single_source_theta_single_ansatz}, but with a different representation of the identity operator. This scheme is motivated by the observed spectral behavior of $\Theta^{fg}G_{gf}$.
Specifically, we replace the products $\Theta^{fg}G_{gf}$ that arise from the $\frac{1}{2}$ terms of $\Theta^{fg}(\mathcal{K}_i-\frac{1}{2})_{gf}$, with the identity matrix, denoted by $\mathcal{I}^f_{\ f}$. In the RWG-BC case, the discretization of the inverse mixed Gram matrix $G^{fg}$ makes the identity operator be represented by $G^{fg}G_{gf}$, which is the identity matrix by definition. Instead, as we mentioned above, $\Theta^{fg}G_{gf}$ only approximates the identity, with a spectrum that depends on the chosen auxiliary space. 
The resulting discretization is given by  
\begin{equation}\label{single_source_identity}
    \begin{split}
     &\bigg(\Big( 
    \Theta^{fg}(\mathcal{K}_0)_{gf}-\frac{1}{2}\mathcal{I}^f_{\ f}\Big)
    \Big(
    -\Theta^{fg}(\mathcal{K}_1)_{fg}+\frac{1}{2}\mathcal{I}^f_{\ f}\Big)
    +\\
    &+\frac{\eta_0}{\eta_1}
    \Theta^{fg}(\mathcal{T}_0)_{gg}
    (-{\Theta^{fg}}^\dagger)
    (\mathcal{T}_1)_{ff}
    \bigg)
    \mathbf{p}^f
    =\Theta^{fg} (\mathbf{\hat{n}} \times \mathbf{e}^{\mathrm{inc}})_g,\\
    \end{split}
\end{equation}
and the physical traces can be obtained by applying \eqref{traces_theta_single_ansatz}.

The motivation for this ad hoc discretization is that $\Theta^{fg}G_{gf}$ acts as a spectral filter, with small eigenvalues corresponding to modes that are poorly represented. We will show in the numerical results that this representation can deteriorate the condition number of the system for some choices of the auxiliary space. The proposed substitution $\Theta^{fg}G_{gf} \to \mathcal{I}^f_{\ f}$ can be interpreted as removing this spurious spectral filtering, replacing it with a uniform action on all components. This is expected to mitigate the degradation of the condition number. 

Note that, in the physical traces representation, the electric trace is given by $(\mathbf{e}\times\mathbf{\hat{n}})^f=\Theta^{fg}\Big(-\mathcal{K}_1+\frac{1}{2}\Big)_{gf}\mathbf{p}^f$, with the identity still represented by $\Theta^{fg}G_{gf}$. In the discretization of the system we apply the substitution $\Theta^{fg}G_{gf} \to \mathcal{I}^f_{\ f}$ to improve the condition number, whereas in the trace evaluation the identity is represented consistently with the discretization of the integral operators, since this choice is observed to provide the most accurate results. 

A double ansatz can be used to obtain both traces in the image of $\Theta^{fg}$. In this case, we solve
\begin{equation}\label{single_source_double_identity}
    \begin{split}
     &\bigg(\Big( 
    \Theta^{fg}(\mathcal{K}_0)_{gf}-\frac{1}{2}\mathcal{I}^f_{\ f}\Big)
    \Big(
    -\Theta^{fg}(\mathcal{K}_1)_{fg}+\frac{1}{2}\mathcal{I}^f_{\ f}\Big)
    +\\
    &+\frac{\eta_0}{\eta_1}
    \Theta^{fg}(\mathcal{T}_0)_{gg}
    (-{\Theta^{fg}}^\dagger)
    (\mathcal{T}_1)_{ff}
    \bigg)
    \mathbf{p}^f
    =\Theta^{fg} (\mathbf{\hat{n}} \times \mathbf{e}^{\mathrm{inc}})_g,\\
    &\bigg(\Big( 
    \Theta^{fg}(\mathcal{K}_0)_{gf}-\frac{1}{2}\mathcal{I}^f_{\ f}\Big)
    \Big(
    -\Theta^{fg}(\mathcal{K}_1)_{fg}+\frac{1}{2}\mathcal{I}^f_{\ f}\Big)
    +\\
    &+\frac{\eta_1}{\eta_0}
    \Theta^{fg}(\mathcal{T}_0)_{gg}
    (-{\Theta^{fg}}^\dagger)
    (\mathcal{T}_1)_{ff}
    \bigg)
    \mathbf{q}^f
    = \Theta^{fg}(\mathbf{h}^{\mathrm{inc}} \times \mathbf{\hat{n}})_g,
    \end{split}
\end{equation}
and the physical traces are obtained by applying \eqref{traces_theta}.

With the proposed scheme in \eqref{single_source_identity} and \eqref{single_source_double_identity}, one can choose an auxiliary space smaller in dimension than the RWG space on the barycentric refinement. For example, the RWG space itself or the BDM space can be chosen as auxiliary spaces. This permits a computationally cheaper discretization of all the integral operators, while maintaining favorable conditioning. On the other hand, the different representations of the identity and the integral operators might introduce consistency errors. We cannot assess the magnitude of these errors a priori, but we will see in the numerical results that the schemes \eqref{single_source_identity}-\eqref{traces_theta_single_ansatz} and \eqref{single_source_double_identity}-\eqref{traces_theta} give remarkably accurate results at reduced computational cost, suggesting that the substitution mainly affects the spectral properties of the system, while the introduced error remains limited.  

\section{Numerical results}\label{sec:numericalexp}

In the numerical experiments, we model the scattering of an incident plane wave propagating along $\mathrm{\hat{z}}$ and polarized along $\mathrm{\hat{x}}$, impinging on a scatterer with relative parameters $\epsilon_r = 2.4$ and $\mu_r = 1.0$, at a fixed wavenumber $\kappa_0=1.0\mathrm{m}^{-1}$, corresponding to a frequency of $47.7\mathrm{MHz}$.
The boundary of the geometry is discretized using flat simplicial triangular elements. 

We present the results for a cuboid-shaped scatterer of dimensions $1.0\mathrm{m} \times 1.0\mathrm{m} \times 0.5\mathrm{m}$, and a sphere of unit radius $R=1.0\mathrm{m}$.

The single-source equation is solved for different choices of the spaces and of the ansatz. In particular we study the following discretizations: 
\begin{itemize}
    \item the discretizations \eqref{single_source_theta_single_ansatz}, \eqref{single_source_theta} with $\Theta^{fg}$, using RWG as the primal space, different choices of the auxiliary spaces,
    \item the discretizations \eqref{discrete_signlesoure}, \eqref{double_discrete_singlesource} with the usual inverse of the mixed Gram matrix $G^{fg}$, using RWG and BC as primal and dual spaces,
    \item the discretizations \eqref{single_source_identity}, \eqref{single_source_double_identity} with $\Theta^{fg}$, using RWG as both primal and auxiliary spaces,
    \item the discretizations \eqref{single_source_theta_single_ansatz}, \eqref{single_source_theta} with $\Theta^{fg}$, using GWP(1) and GWP(2) as primal higher-order spaces, together with GWP(3) and GWP(4) as respective higher-order auxiliary spaces.   
\end{itemize} 
In the first case, where we study the discretization with $\Theta^{fg}$ and the RWG space as the primal space, we limit our analysis to the conditioning properties of the system in the single-ansatz formulation for the cuboid geometry. 
For all the other choices, we also study the accuracy of the discretization, comparing the errors obtained with the single ansatz, where the magnetic trace is expanded in the dual (auxiliary) space, with the errors given by the double ansatz, where both traces are expanded in the primal space.

In order to test the accuracy of each method, we report the relative errors obtained for different quantities. The errors are computed with respect to a reference solution. For the cuboid, the reference solution is obtained by solving the PMCHWT equation for on a fine reference mesh. In particular, for the single-source experiments we explored the mesh-size interval $[0.054\mathrm{m},0.4\mathrm{m}]$, whereas the mesh for the PMCHWT reference solution has $h_{\mathrm{ref}}=\mathrm{0.0176m}$. For the sphere, the reference solution is obtained using the Mie series.

We report the relative errors in the near field by computing the norm of the difference between the electric fields $|\mathbf{E}-\mathbf{E}_{\mathrm{ref}}|$ along the line $y=0.5\mathrm{m},\  z=0.25\mathrm{m}$ for the cuboid, and along the line $y=0.0\mathrm{m},\  z=-0.775\mathrm{m}$ for the sphere, and the relative errors in the bistatic radar cross section for angles $\theta \in [0, 360^\circ]$ on the x-z plane for both geometries. The electric near field is computed as the sum of two contributions: the first is obtained by applying the single-layer potential to the equivalent electric current (i.e., the magnetic trace, expanded in the primal or dual space depending on the ansatz), and the second by applying the double-layer potential to the equivalent magnetic current (i.e., the electric trace, always expanded in the primal space). More precisely, according to the representation theorem, the exterior electric field is obtained by evaluating these potentials with the exterior wavenumber $\kappa_0$ and adding the incident field. Instead, the interior electric field is obtained by evaluating the same potentials with the interior wavenumber $\kappa_1$, using the same currents with opposite sign, and without adding the incident field.  

For the cuboid, we are also able to compute the error on each physical trace separately. Such errors are obtained by computing the norm of the trace with respect to the same trace given by the PMCHWT reference solution. For these global norm errors, we use the following norm equivalent to the energy norm on $H^{-1/2}(\operatorname{div}; \Gamma)$
\begin{equation}
\begin{split}
    &\mathrm{err}_\mathrm{div,-1/2}(\mathbf{j}_h)=\bigg(\int_{\Gamma}\int_{\Gamma}dS'dS \\
    &\Big(
    \kappa \frac{(\mathbf{j}_h(\mathbf{r})-\mathbf{j}_{\mathrm{ref}}(\mathbf{r})) 
    e^{-\kappa R} 
    (\mathbf{\bar{j}}_h(\mathbf{r'})-\mathbf{\bar{j}}_{\mathrm{ref}}(\mathbf{r'}))}{4\pi R}
    + \\
    &+\frac{\nabla\!\cdot\!(\mathbf{j}_h(\mathbf{r})-\mathbf{j}_{\mathrm{ref}}(\mathbf{r})) 
    e^{-\kappa R} 
    \nabla'\!\cdot\!(\mathbf{\bar{j}}_h(\mathbf{r'})-\mathbf{\bar{j}}_{\mathrm{ref}}(\mathbf{r'}))}{4\pi \kappa R}
    \Big)\!\bigg)^{\frac{1}{2}}.
\end{split}
\end{equation}

\subsection{Scattering by the Cuboid}

\paragraph{Discretization with $\Theta^{fg}$. RWG as Primal Space. RWG, BDM, GWP(1), GWP(2) as Auxiliary space\label{sec:auxiliary_lowest_order}}
We report here the results obtained with the single-source equation discretized with $\Theta^{fg}$, using lowest-order RWG primal basis functions, and different choices of auxiliary space, as alternative discretizations to the RWG-BC choice. In this paragraph, we restrict our discussion to the single-ansatz formulation, where we solve \eqref{single_source_theta_single_ansatz}-\eqref{traces_theta_single_ansatz}.

The first case we consider is when the auxiliary space is the RWG space itself.
This configuration was also studied for the preconditioned EFIE in \cite{christiansen2002}. 
This case is interesting since no mesh refinement or higher-order basis functions are employed, and thus reduces the assembly cost of the integral operators. Indeed, all the operators are discretized using only RWG functions in this case: $\mathcal{O}_{gf}=\mathcal{O}_{ff}$. In this discretization, $\Theta^{ff}$, $G_{ff}$, ${\Theta^{ff}}G_{ff}$, and the LHS of \eqref{single_source_theta_single_ansatz}, are all singular matrices, with a kernel of dimension $N_{\Sigma,f}-N_{\Lambda,f}$, with $N_{\Sigma,f},\ N_{\Lambda,f}$ the dimension of the star and the local loop subspaces.

The presence of this kernel implies that the resulting solution lies in a subspace of the original RWG space. 

\begin{figure}
    \centering
        \includegraphics[width=\linewidth]{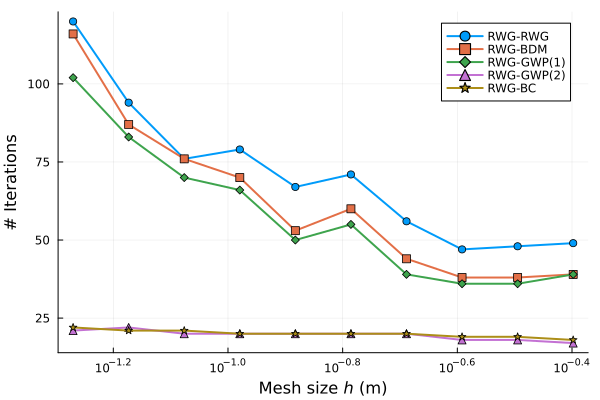}
        \caption{Iterations count vs. mesh size for the cuboid using RWG as primal space, and different choices of the auxiliary space, to reach a residual smaller than $10^{-8}$. Only the single-ansatz case is reported for each choice of the spaces. Smaller auxiliary spaces are affected by a mild ill-conditioning. \label{iters_rwg}}
\end{figure}

Despite the remarkable advantage of the computationally cheaper discretization of the integral operators, the single-source systems discretized in this manner show a mild ill-conditioned behavior. Figure~\ref{iters_rwg} reports the iteration count vs. the mesh size for the cuboid geometry. The number of iterations is higher than for typical well-conditioned single-source systems, showing a mild increase as the mesh is refined. This behavior is not caused by the kernel but by residual ill-conditioning in the image of ${\Theta^{ff}}G_{ff}$ (see Table \ref{tab:eigen_theta_gram} and Table \ref{tab:condition_lhs}). This growth in iteration count represents the main limitation of this configuration making the advantage of the cheaper discretization of the integral operators effective only up to a certain size of the system. 

When choosing BDM space as auxiliary space the system is not singular anymore. This choice was also explored in \cite{christiansen2002} for the EFIE. The small dimension of the BDM space makes this choice still interesting to offset the computational costs of the integral operators discretization. Yet a similar ill-conditioning is observed for ${\Theta^{fg}}G_{gf}$ and for the entire single-source system, limiting the advantage of this choice. A similar mild ill-conditioning happens also when choosing GWP(1) as auxiliary space, as it can be seen from Fig. \ref{iters_rwg}.

When choosing GWP(2) as auxiliary functions, ${\Theta^{fg}}G_{gf}$ becomes well-conditioned, and so does the LHS of \eqref{single_source_theta_single_ansatz}. This observation is consistent with the results reported in \cite{Bourhis2024higher} and \cite{Bouhris2025} with the alternative $\tilde{\Theta}^{fg}$ defined in \eqref{Bourhis_theta}. 
On the other hand, this discretization requires assembling the integral operators with an auxiliary space whose size is comparable to that of the RWG space defined on the barycentric refinement. Consequently, the computational cost of this discretization becomes similar to that of the RWG-BC discretization.

The residual ill-conditioning observed when choosing small auxiliary spaces motivates the discretization in \eqref{single_source_identity} and \eqref{single_source_double_identity}. In the next paragraphs, we focus our error analysis on: the RWG-BC discretization with the usual mixed Gram matrix $G^{fg}$, the discretizations \eqref{single_source_identity} and \eqref{single_source_double_identity} using only RWG functions, and the higher order discretization with $\Theta^{fg}$.

\begin{table}[htbp]
\caption{Condition number of the LHS of the discretized single-source system \eqref{single_source_theta_single_ansatz}, for the cuboid geometry. The condition number reported for the RWG-RWG case is obtained by taking the ratio of the first and the last non-zero eigenvalues.}
\label{tab:condition_lhs}
\centering

\begin{tabular}{|c|cccc|}
\hline
$h$ 
& {RWG-RWG} 
& {RWG-BDM} 
& {RWG-GWP(1)} 
& {RWG-GWP(2)} \\
\hline

$0.4$ & $162$ & $174$ & $183$ & $7.9$ \\
\hline

$0.2$ & $85$ & $78$ & $65$ & $8.9$ \\
\hline

$0.1$ & $197$ & $350$ & $355$ & $9.8$ \\
\hline

$0.05$ & $522$ & $1484$ & $1498$ & $10.5$ \\
\hline

\end{tabular}

\end{table}

\begin{figure*}[!t]
    \centering
    \begin{subfigure}{0.5\textwidth}
        \centering
        \includegraphics[width=\linewidth]{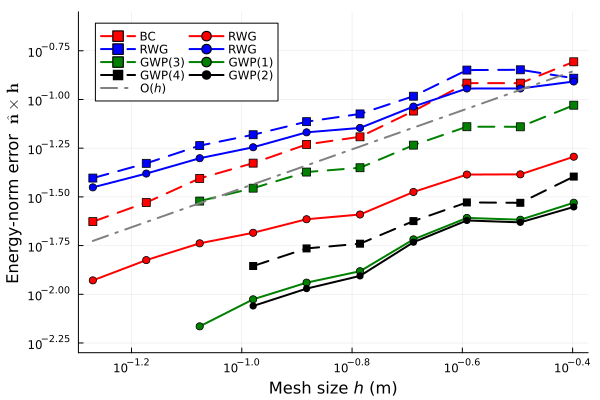}
        \caption{Energy-norm relative errors of the magnetic trace.}
        
    \end{subfigure}\hfill
    \begin{subfigure}{0.5\textwidth}
        \centering
        \includegraphics[width=\linewidth]{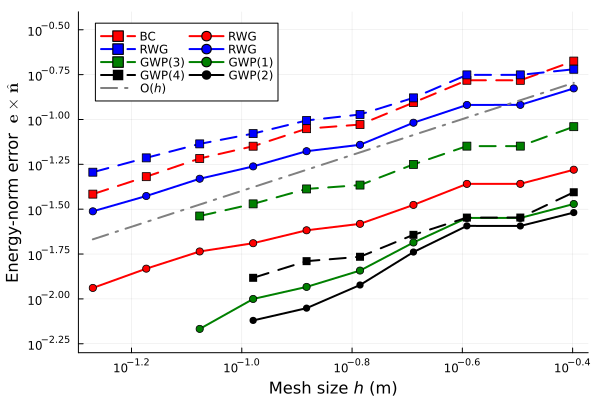}
        \caption{Energy-norm relative errors of the electric trace.}
    \end{subfigure}
    \caption{Energy-norm relative errors of the magnetic trace (left) and the electric trace (right) for the cuboid geometry as a function of the mesh size. Colors denote discretizations: red--RWG-BC with $G^{fg}$, blue--RWG-RWG-Id with $\Theta^{ff}$ and the identity matrix, green--GWP(1)-GWP(3) with $\Theta^{fg}$, black--GWP(2)-GWP(4) with $\Theta^{fg}$. Dashed, square-marked data indicate expansions in the dual (auxiliary) space. Solid, circle-marked data indicate expansions in the primal spaces. 
    The expansion in dual (auxiliary) spaces is affected by larger errors than that in primal spaces. The expansion in higher-order primal spaces further improves the accuracy. }
    \label{fig:energy_norm_errs_ho_and_bc_rwg}
\end{figure*}
\begin{figure*}[!t]
    \centering
    \begin{subfigure}{0.5\textwidth}
        \centering
        \includegraphics[width=\linewidth]{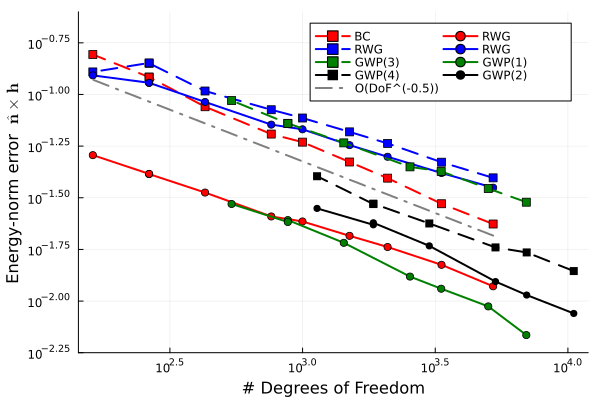}
        \caption{Energy-norm relative errors of the magnetic trace.}
        
    \end{subfigure}\hfill
    \begin{subfigure}{0.5\textwidth}
        \centering
        \includegraphics[width=\linewidth]{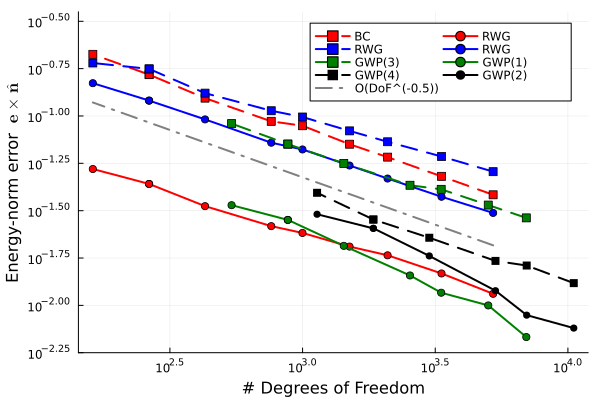}
        \caption{Energy-norm relative errors of the electric trace.}
    \end{subfigure}
    \caption{Energy-norm relative errors of the magnetic trace (left) and the electric trace (right) for the cuboid geometry, as a function of the number of degrees of freedom of the system. Same color and marker conventions as in Fig.~\ref{fig:energy_norm_errs_ho_and_bc_rwg}.
    The expansion in higher-order primal spaces shows a slightly improved convergence slope.}
    \label{fig:energy_norm_dof}
\end{figure*}

\paragraph{RWG as Primal Space. BC as Dual Space}

We consider here the discretization obtained by choosing RWG and BC functions as primal and dual spaces, respectively. We compare the solution accuracy obtained with the single ansatz, in which we solve \eqref{discrete_signlesoure}-\eqref{physical_bc}, with that obtained with the double ansatz, in which we solve \eqref{double_discrete_singlesource}-\eqref{traces_magnetic_double_bc}.
Moreover, by solving the double ansatz system \eqref{double_discrete_singlesource}, we are able to obtain both traces expanded in the RWG space, as in \eqref{traces_magnetic_double_bc}, but we are also able to expand both traces in the BC space by computing 
\begin{equation}
    \label{traces_double_dual}
    \begin{split}
    (\mathbf{\hat{n}} \times \mathbf{h})^g &= \frac{1}{\eta_1}{G^{fg}}^\dagger\big(\mathcal{T}_1)_{ff}\mathbf{p}^f,\\
    (\mathbf{e}\times\mathbf{\hat{n}})^g &=-\eta_1{G^{fg}}^\dagger\big(\mathcal{T}_1)_{ff}\mathbf{q}^f.
    \end{split}
\end{equation}
In this way, we can compare, the accuracy yielded by primal and dual basis expansions for both quantities.
In Fig.~\ref{fig:energy_norm_errs_ho_and_bc_rwg} and Fig.~\ref{fig:energy_norm_dof} we show the scaling of the relative errors in the energy norm with respect to the mesh size and to the degrees of freedom of the system, with the latter corresponding to the number of primal functions of the discretization.
The numerical results in Fig.~\ref{fig:energy_norm_errs_ho_and_bc_rwg}-\ref{fig:energy_norm_dof} show that the expansion in the RWG primal (red circle-marked solid lines) systematically improves the accuracy 
of the corresponding quantity with respect to the expansion in the BC dual space (red square-marked dashed lines). No difference in the asymptotic scaling has been observed, corresponding to $O(h)$, for both RWG and BC expansions. 

\begin{figure}
    \centering
    \includegraphics[width=1.0\linewidth]{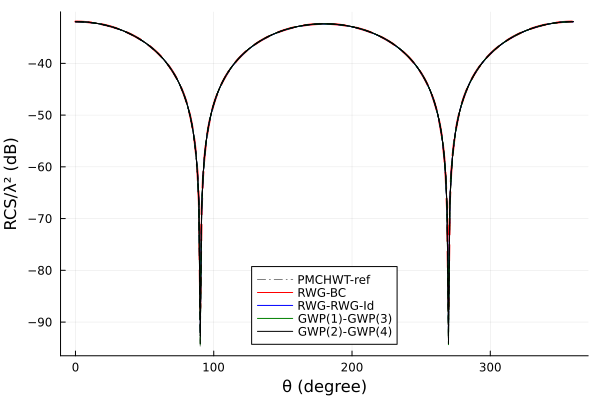}
    \caption{Bistatic radar cross section of the cuboid geometry.}
    \label{fig:far_field_All}
\end{figure}

\begin{figure}
    \centering
    \includegraphics[width=1.0\linewidth]{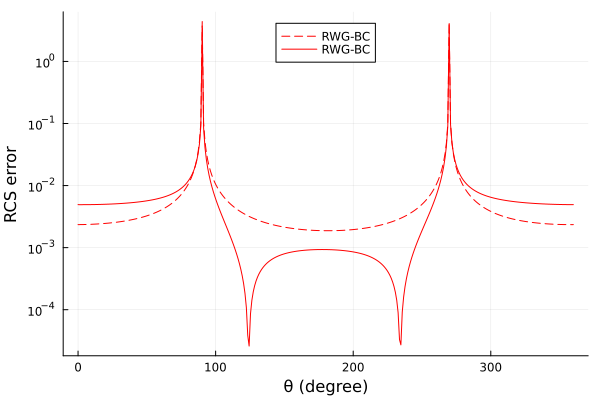}
    \caption{Relative errors in the bistatic radar cross section for the cuboid geometry with the RWG-BC discretization, at mesh size $h=0.054\mathrm{m}$ and $5226$ DoF. Dashed curves: single ansatz. Solid curves: double ansatz. The double-ansatz errors are smaller for angles around $\theta\approx180^\circ$.}
    \label{fig:far_field_errs_rwg_bc}
\end{figure}

For the near-field and far-field errors, we compare the single- and double-ansatz formulations. In Fig.~\ref{fig:nearfield_errs_BC_RT_cuboid} we show the relative near-field errors. Even though the energy norm plots show a clear advantage of using only primal-space expansions, the near-field plots do not show obvious qualitative difference. 
Instead, Fig.~\ref{fig:nearfield_errs_BC_RT_cuboid} shows that the errors are of the same order in the internal region, whereas the single-ansatz shows smaller error in the external region. 
In Fig.~\ref{fig:far_field_All} we show the curve of the bistatic radar cross section, whereas in Fig.~\ref{fig:far_field_errs_rwg_bc} we show its relative error.  
One can see that for observation angles around $\theta \approx 180^\circ$, the double ansatz yields improved accuracy, whereas slightly larger errors are observed near $\theta \approx0^\circ$. 
\begin{figure}
    \centering
        \centering
        \includegraphics[width=\linewidth]{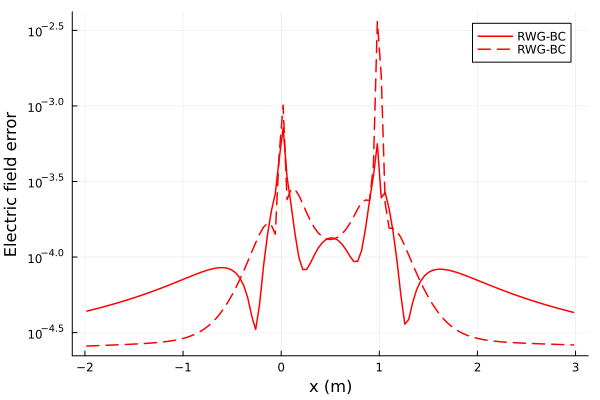}
        \caption{Relative errors in the near field for the cuboid geometry with the RWG-BC discretization, along the line $y=0.5\mathrm{m}$, $z=0.25\mathrm{m}$, at mesh size $h=0.054\mathrm{m}$ and $5226$ DoF. Dashed curves correspond to the single ansatz. Solid correspond to the double ansatz.
        The spikes correspond to the intersections with the surface of the cuboid, where quadrature errors are dominant. The errors are of the same order in the internal region and in proximity of the surface, whereas the single-ansatz results show smaller error in the external region.}
    \label{fig:nearfield_errs_BC_RT_cuboid}
    
    \label{fig:nearfield_errs_BC_RT_cuboid}
\end{figure}

\begin{figure}
    \centering
        \includegraphics[width=\linewidth]{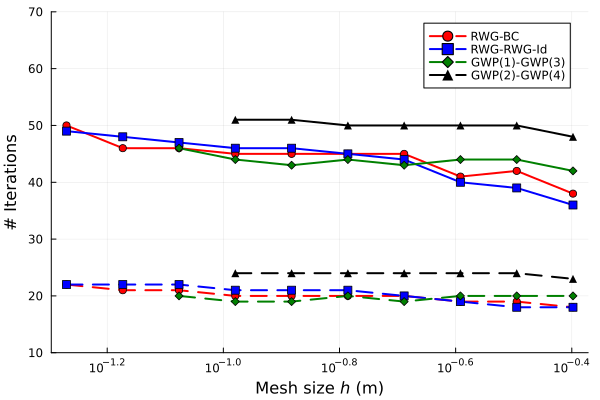}
        \caption{Iterations count vs. mesh size for the cuboid, with lowest-order and higher-order discretizations, to reach a residual smaller than $10^{-8}$. Dashed lines correspond to the single-ansatz formulation. Solid lines correspond to the sum of the iterations of the two solvers in the double ansatz.\label{iters_well_conditioned}}
\end{figure}
\begin{figure}
    \centering
        \includegraphics[width=\linewidth]{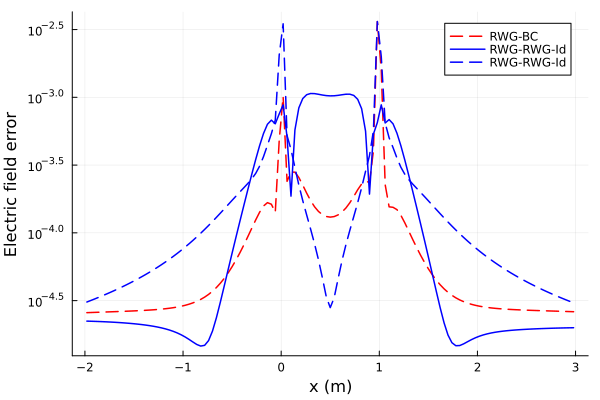}
        \caption{Relative errors in the near field for the cuboid geometry with the RWG-RWG-Id discretization, along the line $y=0.5\mathrm{m}$, $z=0.25\mathrm{m}$, at $h=0.054\mathrm{m}$ and $5226$ DoF. Dashed curves: single ansatz. Solid curves: double ansatz. The single ansatz yields smaller errors in the internal region, whereas the double ansatz yields smaller errors in the external region. Compared to the RWG-BC case, the RWG-RWG-Id discretization yields slightly higher accuracy in either the internal or the external region, depending on the chosen ansatz.}
    \label{fig:nearfield_errs_rwg_rwg_id}
\end{figure}
\begin{figure}
    \centering
    \includegraphics[width=1.0\linewidth]{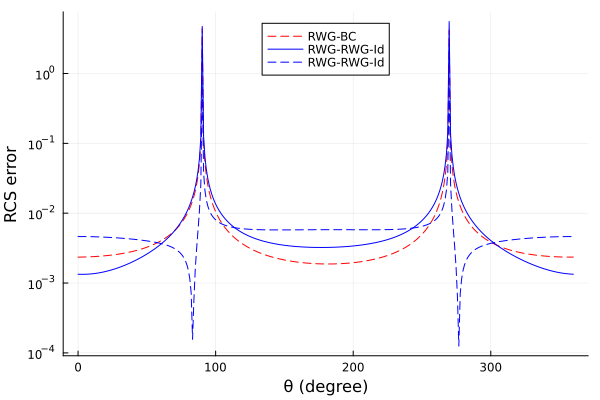}
    \caption{Relative errors in the bistatic radar cross section for the cuboid geometry with the RWG-RWG-Id discretization, at mesh size $h=0.054\mathrm{m}$ and $5226$ DoF. Dashed curves: single ansatz. Solid curves: double ansatz. The double-ansatz errors in the RWG-RWG-Id discretization have a similar magnitude to those in the RWG-BC discretization.}
    \label{fig:far_field_errs_rwg_rwg_id}
\end{figure}
\paragraph{Discretization with $\Theta^{fg}$ and the Identity Matrix. RWG as both Primal and Auxiliary Spaces}

We investigate the solution obtained with the discretization \eqref{single_source_identity} and  \eqref{single_source_double_identity} using RWG as both primal and auxiliary spaces. All operators are discretized using only RWG functions: $\mathcal{O}_{gf}=\mathcal{O}_{ff}$.   

The primary motivation for this strategy is the reduction of the discretization cost associated with the integral operators. In the conforming RWG-BC scheme \eqref{discrete_signlesoure}, the single-layer operator discretized on the barycentric refinement requires the computation of $36N_e^2$ interaction integrals, whereas the double-layer operators discretized with the mixed scheme require $12N_e^2$ interaction integrals, where $N_e$ denotes the number of mesh edges. In the proposed scheme, only $N_e^2$ integrals are required for each operator, as only RWG functions are involved. This results in a substantial reduction in the assembly cost. Other cheap integration strategies, such as the bi-parametric approach, are not applicable in the single-source context, since the regularizing operators are an intrinsic part of the equation and not simply a preconditioner.

We note that, when using fast multipole methods (FMM) or H-matrix methods driven by e.g. addaptive cross approximation (ACA) to suppress memory cost and computational complexity, the number of integrals that must be evaluated explicitly is reduced to roughly $O(N)$, which diminishes the relative impact of the proposed scheme. Nevertheless, the near-field double surface integrals remain a computationally expensive operation.  

Fig.~\ref{iters_well_conditioned} shows that the iterative behavior remains stable, indicating that the favorable conditioning of the system is essentially preserved thanks to the representation of the identity operators with the identity matrix.

In the single ansatz, the physical traces are obtained with 
\begin{equation}\label{traces_theta_single_ansatz_lowest_id}
    \begin{split}
        (\mathbf{\hat{n}} \times \mathbf{h})^f&=\frac{1}{\eta_1}{\Theta^{ff}}^\dagger(\mathcal{T}_1)_{ff}\mathbf{p}^f,\\
        (\mathbf{e}\times\mathbf{\hat{n}})^f&=\Theta^{ff}\Big(-\mathcal{K}_1+\frac{1}{2}\Big)_{ff}\mathbf{p}^f,
    \end{split}
\end{equation}
whereas in the double ansatz, we get
\begin{equation}\label{traces_theta_double_ansatz_lowest_id}
    \begin{split}
        (\mathbf{\hat{n}} \times \mathbf{h})^f&=\Theta^{ff}\Big(-\mathcal{K}_1+\frac{1}{2}\Big)_{ff}\mathbf{q}^f,\\
        (\mathbf{e}\times\mathbf{\hat{n}})^f&=\Theta^{ff}\Big(-\mathcal{K}_1+\frac{1}{2}\Big)_{ff}\mathbf{p}^f.
    \end{split}
\end{equation}
Although the magnetic trace is expanded in the RWG space in both ans\"atze, the two representations are different as one is an expansion in the image of ${\Theta^{ff}}^\dagger$ and the other is an expansion in the image of $\Theta^{ff}$.

The energy norm errors in Fig.~\ref{fig:energy_norm_errs_ho_and_bc_rwg}-\ref{fig:energy_norm_dof} show that the quantities expanded in the image of ${\Theta^{ff}}^\dagger$ (blue square-marked dashed lines) are affected by slightly larger errors than those expanded in the image of $\Theta^{ff}$ (blue circle-marked solid lines). From Fig.~\ref{fig:nearfield_errs_rwg_rwg_id}, the single-ansatz formulation yields better accuracy in the near field in the internal region, whereas the double-ansatz formulation yields better accuracy in the external region. The far-field accuracy in the bistatic radar cross section is slightly better in the double-ansatz case (Fig.~\ref{fig:far_field_errs_rwg_rwg_id}). Compared to the RWG-BC discretization at the same mesh sizes, the energy-norm errors are larger, but the near-field errors are smaller in either the external or the internal region, depending on the chosen ansatz. The far-field accuracy in the double-ansatz formulation is close to that of the RWG-BC discretization.

The results suggest that the representation of the identity operators with the identity matrix effectively preserves the conditioning of the system, whereas the consistency error introduced is significant only in the energy norm of the traces. By choosing the double-ansatz formulation such an error does not significantly affect the external near and far field. 

\begin{figure}
    \centering
        \includegraphics[width=\linewidth]{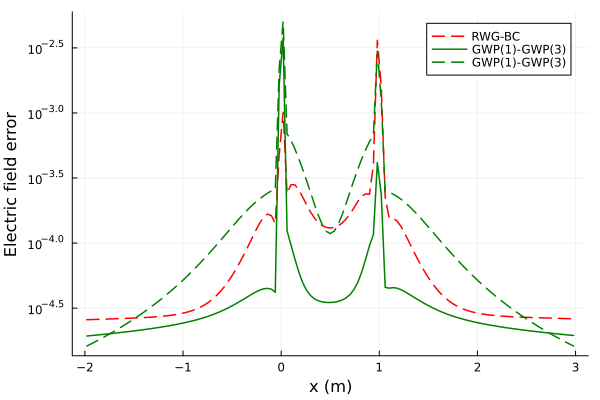}
    \caption{Relative errors in the near field for the cuboid geometry with the GWP(1)-GWP(3) discretization, at $h=0.105\mathrm{m}$ and $5010$ DoF in the primal space. Dashed curves: single ansatz. Solid curves: double ansatz. The double ansatz yields better accuracy than the single ansatz in proximity of the scatterer. Compared to the lowest order RWG-BC results with similar $5226$ DoF, the double-ansatz formulation yields similar accuracy in the external region, and higher accuracy in the internal region.}
    \label{fig:nearfield_errs_GWP1_GWP3_cuboid}
\end{figure}
\begin{figure}
        \centering
        \includegraphics[width=\linewidth]{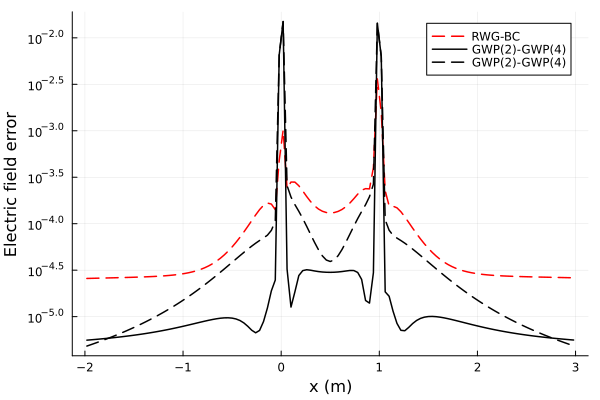}
    \caption{Relative errors in the near field for the cuboid geometry with the GWP(2)-GWP(4) discretization, at $h=0.164\mathrm{m}$ and $5334$ DoF in the primal space. Dashed curves: single ansatz. Solid curves: double ansatz.
    The double ansatz yields better accuracy in proximity of the scatterer. Compared to the lowest order RWG-BC results with similar $5226$ DoF, the GWP(2)-GWP(4) discretization yields better accuracy per degree of freedom in both ans\"atze.}
    \label{fig:nearfield_errs_GWP2_GWP4_cuboid}
\end{figure}
\begin{figure}
    \centering
    \includegraphics[width=1.0\linewidth]{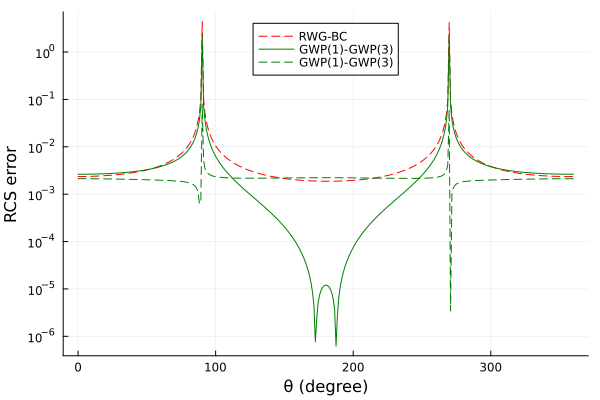}
    \caption{Relative errors of the bistatic radar cross section for the cuboid geometry with the GWP(1)-GWP(3) discretization, at $h=0.105\mathrm{m}$ and $5010$ DoF. Dashed curves: single ansatz. Solid curves: double ansatz. The double-ansatz errors are smaller for angles around $\theta\approx180^\circ$. Compared to the lowest order RWG-BC results with similar $5226$ DoF, the double ansatz yields better accuracy for angles around $\theta\approx180^\circ$, whereas the single-ansatz yields similar accuracy.}
    \label{fig:far_field_errs_gwp1_primal_cuboid}
\end{figure}

\begin{figure}
    \centering
    \includegraphics[width=1.0\linewidth]{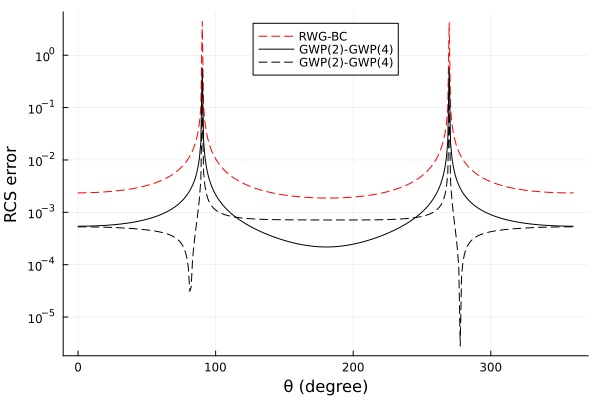}
    \caption{Relative errors of the bistatic radar cross section for the cuboid geometry with the GWP(2)-GWP(4) discretization, at $h=0.164\mathrm{m}$ and $5334$ DoF. Dashed curves: single ansatz. Solid curves: double ansatz. The double-ansatz errors are slightly smaller for angles around $\theta\approx180^\circ$. Compared to the lowest order RWG-BC results with similar $5226$ DoF, both the ans\"atze yield better accuracy per degree of freedom.}
    \label{fig:far_field_errs_gwp2_primal_cuboid}
\end{figure}

\paragraph{Higher-order Discretization with $\Theta^{fg}$. GWP(1), GWP(2) as Primal Spaces. GWP(3), GWP(4) as Auxiliary Spaces}
We next consider the case in which higher-order GWP spaces are used both as primal and auxiliary space. In particular we explored the cases where the primal space is GWP($i$) with $i=1,2$, and the auxiliary space is GWP$(i+2)$. In the single-ansatz formulation we solve \eqref{single_source_theta_single_ansatz}-\eqref{traces_theta_single_ansatz}. In the double-ansatz formulation, we solve \eqref{single_source_theta}-\eqref{traces_theta}.

In Fig. \ref{iters_well_conditioned} we show the number of iterations w.r.t. the mesh size for the cuboid case. We observed a well-conditioned behavior for these choices, similarly to the observation made in \cite{Bourhis2024higher} and \cite{Bouhris2025}.
In contrast to the lowest-order case, the choice GWP($i+2$) instead of GWP($i$) on the barycentric refinement as auxiliary space reduces the total number of degrees of freedom for $i>0$. Indeed, the number of basis functions of the general GWP($i$) space is $N_{i}=(i+1)(\frac{2}{3}i+1)N_e$, where $N_e$ is the number of edges of the mesh. The ratio between the number of functions in the auxiliary and the primal spaces is
\begin{equation}
    \frac{N_{\mathrm{Auxil.}}}{N_{\mathrm{Primal}}}= \frac{N_{i+2}}{N_i}=\frac{(i+3)(2i+7)}{(i+1)(2i+3)},
\end{equation}
which is less than the factor of 6 associated with the barycentric refinement for any $i>0$. In particular, in our case, $\frac{N_{i+2}}{N_{i}}\approx3.6, 2.619$, for $i=1,2$, respectively.  

From Fig.~\ref{fig:energy_norm_errs_ho_and_bc_rwg}, traces expanded in higher-order primal and auxiliary GWP functions exhibit lower errors than lowest-order expansions in the energy norm for all mesh sizes, as expected due to the increased number of basis functions. When plotted against the degrees of freedom, the energy-norm error of higher-order discretizations shows a slightly improved asymptotic convergence behavior for primal-space expansions (Fig.~\ref{fig:energy_norm_dof}), as well as improved accuracy per degree of freedom for expansions in GWP(1) primal functions.
Higher-order primal-space expansions yield systematically improved accuracy in the energy norm than the corresponding higher-order auxiliary-space expansions. 

Regarding near-field accuracy, in both cases, the double-ansatz formulation produces smaller errors than the respective single-ansatz formulations in proximity of the scatterer (Fig.~\ref{fig:nearfield_errs_GWP1_GWP3_cuboid}–\ref{fig:nearfield_errs_GWP2_GWP4_cuboid}). For the far-field results, the bistatic radar cross-section errors (Fig.~\ref{fig:far_field_errs_gwp1_primal_cuboid}–\ref{fig:far_field_errs_gwp2_primal_cuboid}) show improved accuracy in the double ansatz only for observation angles around $\theta \approx 180^\circ$, whereas slightly larger errors are observed near $\theta \approx 0^\circ$, similarly to the lowest-order discretization.

Compared with the RWG-BC discretization, the GWP$(2)$-GWP$(4)$ discretization achieves a clear improvement in near-field and far-field accuracy per degree of freedom in both ans\"atze, whereas the GWP$(1)$-GWP$(3)$ discretization yields improved accuracy in the near field mainly in proximity of the scatterer in the double-ansatz formulation, and in the far field only for observation angles around $\theta \approx 180^\circ$ in the double-ansatz formulation.

\subsection{Scattering by the Sphere}
In the sphere case, no sharp edges are present, and therefore the spurious continuity issues typically associated with the dual space do not arise. Instead, a geometric error is present, due to the approximation of the curved surface by planar triangular meshes. Moreover, an analytic reference solution is available through the Mie series.

\paragraph{RWG as Primal Space. BC as Dual Space}

\begin{figure}
    \centering
        \centering
        \includegraphics[width=\linewidth]{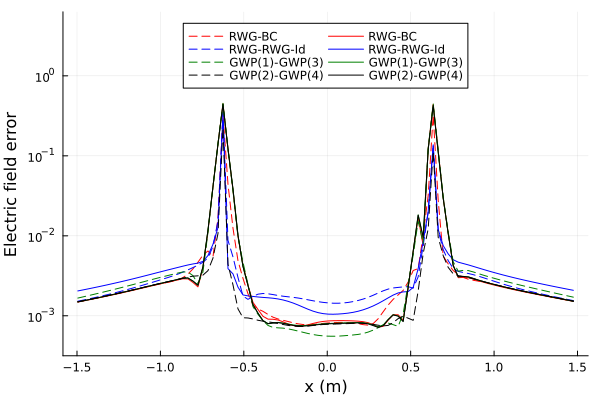}
        \caption{Relative errors in the near field for the sphere, along the line $y=0\mathrm{m}$, $z=-0.775\mathrm{m}$, at mesh size $h=0.164\mathrm{m}$. The spikes correspond to the intersections with the surface of the sphere, where quadrature errors are dominant. Dashed curves: single ansatz. Solid curves: double ansatz. No significant difference in accuracy is observed between the choices of ansatz and discretization.}
    
    \label{fig:nearfield_errs_all_SPHERE}
\end{figure}

\begin{figure}
    \centering
        \centering
        \includegraphics[width=\linewidth]{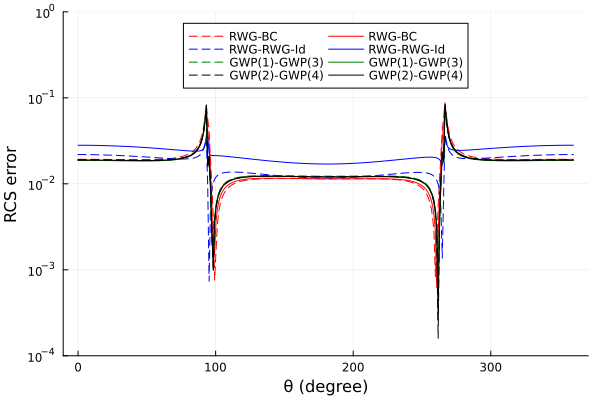}
        \caption{Relative errors in the bistatic radar cross section for the sphere, at $h=0.164\mathrm{m}$. Dashed curves: single ansatz. Solid curves: double ansatz. Solid curves are obtained with the double ansatz. No significant difference in accuracy is observed between the choices of ansatz and discretization. Only the RWG-RWG-Id discretization shows slightly larger errors in the double-ansatz results.}
    
    \label{fig:farfield_errs_all_SPHERE}
\end{figure}

In the sphere case, near-field and far-field results are compared against the Mie series solution. In Fig.~\ref{fig:nearfield_errs_all_SPHERE} 
the relative errors in the near field along a line are shown. In Fig. \ref{fig:farfield_errs_all_SPHERE} we show the errors in the bistatic radar cross section. 
For the RWG-BC discretization, the numerical results obtained with the single and the double ans\"atze are almost identical in both the near-field and the far-field errors.


\paragraph{Discretization with $\Theta^{fg}$ and the Identity Matrix. RWG as both Primal and Auxiliary Spaces}

In the sphere case, the near-field errors of the RWG-RWG-Id discretization do not exhibit any degradation with respect to the RWG-BC one (Fig.~\ref{fig:nearfield_errs_all_SPHERE}). The bistatic radar cross-section errors in Fig.~\ref{fig:farfield_errs_all_SPHERE} do not show any significant degradation for the single-ansatz results, and show slightly larger errors for the double-ansatz results. 
The data show that the ad hoc representation of the identity does not adversely affect the quality of the solution.


\paragraph{Higher-order Discretization with $\Theta^{fg}$. GWP(1), GWP(2) as Primal Spaces. GWP(3), GWP(4) as Auxiliary Spaces}

For the sphere, no significant improvement is observed when increasing the order of the primal space from RWG to GWP$(i)$ at the same mesh size. The accuracy of both formulations is limited by the geometric approximation of the surface using flat triangles. This was observed for both the near-field errors (Fig.~\ref{fig:nearfield_errs_all_SPHERE}) and the bistatic radar cross section errors (Fig.~\ref{fig:farfield_errs_all_SPHERE}).
 
\section{Conclusion}

This work studied the role of primal and dual basis functions in the discretization of the single-source integral equation, with particular emphasis on the impact of dual space expansions on solution accuracy. While the conforming discretization provides a stable and well-conditioned solver, it intrinsically enforces an asymmetric representation where one of the physical traces is expanded in the dual space, which may limit the approximation properties of the formulation. A single-source discretization that allows both traces to be accurately computed in the primal space by solving a single BIE remains elusive.

To address this limitation, we introduced a double-ansatz formulation that allows both the electric and magnetic traces to be consistently expanded in the primal space by solving two single-source equations, which share the same integral operators. Thus, the involved integral operators need to be assembled only once, while two different systems need to be solved. 

Moreover, in this work we employed the $\Theta$ map proposed in \cite{christiansen2002} to generalize the construction of the dual space and extend the single-source to higher-order discretizations. 

The numerical investigations show that, for lowest-order and higher-order discretizations, the influence of the trace expansion space on accuracy is mainly on the energy-norm errors. The near field is not affected in the lowest-order case, and mildly affected in the higher-order discretization. When expanding both traces in the primal space, errors in the bistatic radar cross section of the cuboid geometry show better accuracy only at observation angles around $180^\circ$, for both lowest-order and higher-order discretizations.

Higher-order primal functions achieve better accuracy per degree of freedom in both near-field and far-field quantities, provided that the boundary of the geometry is discretized with sufficient precision. The $\Theta$ map enables the use of higher-order functions also in place of the dual space, further optimizing the number of degrees of freedom.


Lastly, we leveraged the flexibility of $\Theta$ map, together with an ad hoc representation of the identity operator, to obtain a scheme where all the integral operators are discretized using only RWG functions.
With this, we obtained an accuracy close to that obtained with the discretization with RWG and BC functions, but with significantly reduced computational costs in the discretization of the integral operators.

\bibliographystyle{ieeetr}
\bibliography{bibsinglesource}

\end{document}